\documentclass[12pt,preprint]{emulateapj}

\usepackage{epsfig}
\def \farcs{\hbox{$.\!\!^{\prime\prime}$}}

\lefthead{Muzzin et al.}
\righthead{NIR Properties of Moderate-Redshift Galaxy Clusters}
\begin{document}

\title{Near-Infrared Properties of Moderate-Redshift Galaxy Clusters:
  Luminosity Functions and Density Profiles} 

\author{Adam Muzzin, H. K. C. Yee}
\affil{adam.muzzin@utoronto.ca \\ Dept. of Astronomy \& Astrophysics, University of Toronto \\
   50 St. George Street, Toronto, Ontario, Canada, M5S 3H4  } 
\author{Patrick B. Hall\altaffilmark{1}}
\affil{phall@yorku.ca \\ Department of Physics \& Astronomy,
York University \\ 4700 Keele Street, Toronto, Ontario, Canada, M3J 1P3} 
\author{E. Ellingson}
\affil{Erica.Ellingson@colorado.edu \\ Center for Astrophysics and
  Space Astronomy, University of Colorado at Boulder \\ CB389, Boulder, CO, 80309}
\author{H. Lin\altaffilmark{1}}
\affil{hlin@fnal.gov \\ Fermi National Accelerator Laboratory \\ P.O. Box 500, Batavia, IL 60510} 
\altaffiltext{1}{Visiting Astronomer, Kitt Peak National Observatory,
  National Optical Astronomy Observatory, which is operated by the
  Association of Universities for Research in Astronomy, Inc. (AURA)
  under cooperative agreement with the National Science Foundation.}
\begin{abstract}
 We present K-band imaging for 15 of the Canadian Network for
 Observational Cosmology (CNOC1) clusters.  The extensive
 spectroscopic dataset available for these clusters allows us to determine
 the cluster K-band luminosity function and density profile without the
 need for statistical background subtraction.  The luminosity density
 and number density profiles
 can be described by NFW models with concentration parameters of c$_{l}$ =
 4.28 $\pm$ 0.70 and c$_{g}$ = 4.13 $\pm$ 0.57 respectively.  Comparing
 these to
 the dynamical mass analysis of the same clusters shows that the
 galaxy luminosity and number density profiles are similar to the
 dark matter profile, and are not less concentrated like in local clusters. 
 The luminosity functions show that the
 evolution of K$^{*}$ over the redshift range 0.2 $< z <$ 0.5 is
 consistent with a scenario where the majority of stars in cluster
 galaxies form at high-redshift ($z_{f} >$ 1.5) and evolve passively
 thereafter.  The best-fit for the faint-end slope of the luminosity
 function is $\alpha$ = -0.84 $\pm$ 0.08, which indicates that it does
 not evolve between $z =$ 0 and $z$ = 0.3.   
 Using Principal Component Analysis of the spectra we classify cluster
 galaxies as either star-forming/recently-star-forming (EM+BAL) or
 non-starforming (ELL) and compute their respective
 luminosity functions.  The faint-end slope of the ELL
 luminosity function is much shallower than for the EM+BAL galaxies at
 $z = 0.3$,
 and suggests the number of faint ELL galaxies in clusters decreases by a factor
 of $\sim$ 3 from $z = 0$ to $z = 0.3$.   
 The redshift evolution of K$^{*}$ for both EM+BAL and ELL types is 
 consistent with a passively evolving stellar population formed at
 high-redshift.  Passive evolution in both classes, as
 well as the total cluster luminosity function, demonstrates that the bulk
 of the stellar population in all bright cluster galaxies is formed at high-redshift and
 subsequent transformations in morphology/color/spectral-type have little
 effect on the total stellar mass.  
\end{abstract}

\keywords{cosmology: dark matter $-$ large-scale structure of universe
\\ \hspace{15.0cm} galaxies: clusters: photometry $-$ fundamental parameters}

\section{Introduction}
Galaxy clusters are fundamental tools in the study of galaxy
evolution because they are unique locations in the universe, where the high-density environment
produces a population of galaxies that is different from the
general field.  The ability to predict how and when cluster galaxies
are assembled, and their subsequent evolution is an important test for any 
model of galaxy formation.  Unfortunately, the
cluster population transforms significantly in morphology, color, and
star-formation properties over the  
redshift range 0 $< z <$ 0.5 and thus far, our understanding of
this evolution is incomplete. 
At low redshift ($z <$ 0.2), clusters
are primarily composed of a population of quiescent
early-type galaxies which obey a tight color-magnitude relation
(CMR, e.g., Bower et al. 1992; Lopez-Cruz et al. 2004), morphology-density relation
(Dressler 1980; Goto et al. 2003a), and spectral type-density relation
(Tanaka et al. 2004; Gomez et al. 2003; Lewis et
al. 2002). 
\newline\indent
At higher redshifts (0.2 $< z <$ 0.5) the cluster galaxy population is no longer
completely dominated by early-types.  The number of blue
galaxies in clusters increases
(the Butcher-Oemler effect, e.g., Butcher \& Oemler 1984; Ellingson et
al. 2001; Andreon et al. 2004) and the morphological mix of galaxies also changes as the
proportion of spiral galaxies increases at the expense of 
the early-type (primarily S0) population (e.g., Dressler et al. 1997;
Postman et al. 2005; Smith et al. 2005).
Despite these major changes in star-formation properties and morphology, studies of the fundamental plane
(e.g., van Dokkum et al. 1998; van Dokkum \& Stanford 2003; Holden et
al. 2005) and the CMR (e.g., Stanford et al. 1998; Gladders et
al. 1998; Holden et al. 2004) of early-type galaxies (both of which
include S0 galaxies) show that
their stellar populations are extremely old and
consistent with a passively evolving population formed at high
redshift ($z_{f} >$ 2-3).  The differences in the
star-formation history of cluster galaxies and the average age of their
stellar populations could be reconciled by postulating that the
predecessors of low-redshift early-types are high-redshift
late-types which form the bulk of their stars at high-redshift (van
Dokkum et al. 2001).  If
transformations in color and morphology are primarily passive
(i.e., from the truncation of star formation by strangulation; e.g., Balogh
et al. 1999; Abraham et al. 1996; Treu et al. 2003; Goto et al. 2003b) then
they should leave little imprint on the overall stellar population.  The
prediction that the majority of evolution in cluster galaxies from $z
\sim$ 0.5 to the present is 
simply the passive transformation of late-type galaxies into
early-type galaxies is in good qualitative agreement with the data;
however, it is possibly too simplistic a picture. 
\newline\indent
 Recent work 
suggests that galaxy-galaxy mergers may play a significant role in driving cluster
galaxy evolution at high redshift.  Lin et al. (2004, hereafter L04) compared the Halo Occupation
Distribution (HOD, the number of galaxies in a dark matter halo of a
 given mass) of 0.2 $< z <$ 0.9 clusters to that of $z <$ 0.1 clusters and
showed that it is {\it larger} by a factor of $\sim$ 2 in the high-redshift
clusters.  This result implies that numerous mergers or tidal
 disruptions in the cluster
environment at moderate to high redshift are required to reduce the number of bright cluster
galaxies, and match the local HOD.  The prediction of a large merger fraction is consistent with
observations of MS1054-03 at $z$ = 0.83 by Tran et al. (2005) and van Dokkum et
al. (1999) who show that 16\% of the cluster galaxies are currently
undergoing major mergers.  Understanding how the merger rate interplays with the
transformation in color, spectral-type, and morphological-type of the
cluster population without significantly altering the overall
stellar population remains a challenge. 
\newline\indent
The K-band luminosity function (LF) of clusters at this redshift can
provide useful information on this problem. 
K-band light suffers little contamination from recent 
star formation and dust; furthermore, the K-corrections are small, negative,
and nearly independent of galaxy type (e.g., Poggianti 1997; Mannucci
et al. 2001).  These advantageous properties mean that K-band light is
closely related to the total 
stellar mass of a galaxy (Brinchmann \& Ellis 2000; Rix \& Rieke 1993)
and therefore, the K-band LF is nearly 
analogous to the galaxy stellar mass function.  The dependence on the
stellar mass contained in galaxies
makes the K-band LF a useful check on the merger rate, because its evolution will
proceed quite differently depending on the number of major mergers.  Furthermore,
the K-band LF is also sensitive to 
the age of the stellar populations in cluster galaxies because it
puts constraints on 
luminosity evolution of the stellar population.  Most commonly, luminosity
functions are fit to a Schechter (1976) function of the form:
\begin{equation}
\phi(\mbox{K}) = (0.4 \mbox{ln} 10)\phi_{*}(10^{0.4(\mbox{\scriptsize
    K$^{*}$-K})})^{1+\alpha}\mbox{exp}(-10^{0.4(\mbox{\scriptsize $K^{*}$-K})}),
\end{equation}
where $\alpha$ is the faint-end slope; $\phi_{*}$, the normalization;
and K$^{*}$ is the ``characteristic'' magnitude, which indicates the
transition between
the power-law behavior of the faint-end and the
exponential behavior of the bright-end.
\newline\indent
The advent of the 2 Micron All-Sky Survey (2MASS, Skrutskie et al. 2006) has facilitated a
considerable number of studies of the Near Infrared (NIR) LF of
clusters in the local universe (i.e., $z$ $<$ 0.1) using a
very homogeneous set of data.  These studies now provide a crucial
comparison sample for higher-redshift NIR cluster LFs.  Balogh et al. (2001, hereafter B01), Kochanek et
al. (2003), Lin et al. (2003), L04, Rines et
al. (2004, hereafter R04) and Ramella et al. (2004) have
all investigated the K-band LF of local clusters and groups using 2MASS
data.  The LFs vary somewhat from sample to
sample; however, it appears that the overall shape of the K-band LF in local
clusters is similar to the local field.  The only notable
difference being that K$^{*}$ is
slightly ($\sim$ 0.2 - 0.4 magnitudes) brighter in clusters, and
K$^{*}$ in groups is closer to, if not equal, the field value (Ramella
et al. 2004; L04).  
\newline\indent
At higher redshift, the K-band LF of clusters is not as well constrained.
Thus far, the best measurement of the evolution of the cluster K-band
LF comes from the study of de Propris et al. (1999, hereafter dP99).
Using a heterogeneous set of 38 clusters with 0.1 $< z <$ 1.0, they show
that the brightening of K$^{*}$ with increasing redshift is consistent
with a passively evolving stellar population with a formation redshift
($z_{f}$) of $\sim$ 2-3.  Subsequent measurements of the
K-band LF in a handful of $z > 1$ clusters have been
made (Kodama \& Bower 2003; Ellis \& Jones 2004; Toft et
al. 2004; Strazzullo et al. 2006) and those data further confirm the passive evolution
scenario.  The interpretation of these results has been that the majority of the stellar
mass in bright cluster galaxies is already
assembled into IR-bright (and therefore massive) galaxies by $z \sim
1$.  It is difficult to reconcile these data with the prediction of a
significant number of mergers in the cluster environment.  
\newline\indent
We have obtained wide-field
K-band imaging for 15 of the 16 CNOC1 clusters (Yee et al. 1996a) to a depth of $\sim$
K$^{*}$ + 2.  With these data we examine the redshift evolution of the
K-band cluster
LF, the difference between the field and cluster LFs, the LFs of clusters of
different mass, and the K-band luminosity and number density profile of clusters at 0.19
$< z <$ 0.54 using a homogeneous, X-ray-selected cluster sample.  We have IR-imaging and spectroscopy to R $\sim$
1.5R$_{200}$ for most clusters and this radial coverage allows us to compute the LFs for
the entire virialized region without the
need for statistical background subtraction.  Furthermore, the spectroscopy also
allows us to classify the galaxies into star-forming, and
non-star-forming types and determine how the LFs of
these classes differ.
\newline\indent
This paper is the first in a series of three which examine the NIR
properties of moderate redshift clusters.  The HOD, mass-to-light ratios and correlation between the cluster
physical properties and K-band luminosity/richness are presented in
Muzzin et al. (2006, hereafter Paper II).  The 
evolution of the color-magnitude relation and the K-band selected
Butcher-Oemler effect will be presented in a third paper (Yee et
al. in-prep, hereafter Paper III).
\newline\indent
The structure of this paper is as follows:  In $\S$2 we describe our
observations, and in $\S$3 the data reduction and calibration.
Section 4 discusses the weighting scheme based on the spectroscopic catalogue
that is used to construct both the density profiles and luminosity
functions.  In $\S$5 we compute the K-band luminosity and number
density profile of the
clusters and compare with the dark matter profile.  In $\S$6
we determine the cluster K-band LF
for all cluster galaxies as a function of redshift and cluster mass
and compare to field LFs in the literature.   
In $\S$7 we divide the galaxies into two spectral classes and show the
dependence of these LFs on redshift and cluster mass.  We conclude
with a summary in $\S$8.  When
computing magnitudes and angular sizes we adopt an 
$\Omega_{m}$ = 0.3, $\Omega_{\Lambda}$ = 0.7, H$_{0}$ = 70 km s$^{-1}$
  Mpc $^{-1}$ cosmology.  All magnitudes quoted throughout the paper
  are on the Vega system.

\section{Observations}
The CNOC1 clusters are an X-ray selected sample of 16 intermediate (0.19
$< z <$ 0.54) redshift clusters (Yee et al. 1996a), and are likely the most well-studied clusters in
this redshift range.  The masses (Carlberg et al. 1996, Borgani et al. 1999), mass-profiles
(Carlberg et al. 1997a, 1997b; van der Marel et al. 2000), X-ray temperatures
(Mushotzky \& Scharf 1997; Henry 2000; Lewis
et al. 1999; Hicks et al. 2006), X-ray luminosities (Ellis \& Jones
2002), richnesses (Yee \& Ellingson 2003), and 
stellar populations (Abraham et al. 1996; Balogh et al. 1999; Diaferio et al. 2001;
Ellingson et al. 2001) have all been examined in detail.  
For our sample we selected 15 of the 16 CNOC1 clusters for K-band
imaging.  The cluster MS0906+11 was not observed because it is was
shown to be a strong binary in redshift-space by Carlberg et al. (1996) and
therefore the cluster dynamical mass measurement is unreliable.  

\subsection{Optical Photometry and Spectroscopy}
Gunn {\it g} and {\it r} band imaging data were obtained at the 3.6m
Canada-France-Hawaii-Telescope (CFHT) as part of the original CNOC1
project (Yee et al. 1996a)
using the Multi-Object Spectrograph (MOS) camera in imaging mode.
The 5$\sigma$ depth of the optical photometry varies from
cluster-to-cluster but is typically 23.7 - 24.3 mag in $g$, and 23.5 -
24.0 mag in $r$.  The CNOC1 project also acquired $>$ 2500 spectroscopic redshifts in the
fields of the 15 clusters.   The spectroscopy was obtained using 1-10 masks of
$\sim$ 100 slits per cluster.  Of the 2500 redshifts, approximately half are 
cluster members.  The spectroscopic catalogues were chosen as an $r$-band 
magnitude-limited, complete sample, but are sparsely sampled.  The
spectroscopic magnitude limits are $r$
= 20.5, 21.5, and 22.0 for clusters at $z <$ 0.3, 0.3 $< z <$
0.45, and $z > 0.45$ respectively.  Using the passive evolution color-redshift model for
an early-type galaxy from Kodama et al. (1998), these spectroscopic completeness limits correspond to K-band limits of $\sim$
17.0, 17.5, and 18.0 mag for the same redshift bins.  The definition
of the completeness limiting magnitude is that all
galaxies brighter than the limit, {\it that had slits
placed on them}, have reliably determined redshifts.  However, the
sparse sampling means that not all
galaxies in the field brighter than the completeness limit
had slits placed on them.  Therefore, use of the redshift catalogues requires a
selection function that corrects for the sparse sampling of the data
($\S$4).  
\newline\indent
Complete details of the optical observations, reductions, 
photometry and redshift determination of these data can be found in
Yee et al. (1996a).    Our analysis is based on the original
photometric and spectroscopic catalogues (Ellingson et al. 1998, 1997;
Abraham et al. 1998; Yee et al. 1998, 1996b) and additional
unpublished spectroscopy for $z <$ 0.3 clusters.  

\subsection{Near Infrared Observations}
K-band imaging for 14 of the 15 clusters was obtained at the
Kitt Peak National Observatory (KPNO) 2.1m telescope using the Ohio
State / NOAO Infrared Imaging Spectrograph (ONIS).  The ONIS detector
is a 1024 $\times$ 1024 InSb array with 2 working quadrants which makes it
effectively a 1024 $\times$ 512 imager with a pixel scale of 0\farcs288/pixel.
 The observations were made on UT 1999 February
3-4 and 1999 June 3, and were taken in a Mauna Kea 
Observatory filter set version of the K$_s$ filter (Tokunaga et al. 2002), which is 
nearly identical to the 2MASS K$_s$ filter.  Hereafter we treat them 
as identical and refer to the filter simply as the ``K-band''.
All three nights were photometric and standard stars
were observed at different airmasses for photometric calibration.  The seeing in the images
is typically around 1\farcs0, but ranges from 0\farcs8 to 1\farcs2.
Because of time constraints, we were unable to obtain data for one
cluster (MS0440+02) during the February run.  K-band imaging of
MS0440+02 was instead obtained using the PISCES camera on the KPNO
2.5m telescope on UT 1999 February 27 during 
an imaging run for CNOC2 fields.  The PISCES camera is a 1024 $\times$
1024 HAWAII array with a pixel scale of 0\farcs495/pixel.  The PISCES
imaging was done using the K$_{s}$ filter.  Table 1 summarizes our
observational data.


\subsection{Pointing Strategy and Field Coverage}
\indent
The goal of the original CNOC1 observations was to obtain photometry
and spectroscopic redshifts of cluster members beyond the cluster
virial radius (R$_{vir}$), in order to determine the cluster dynamical
mass-to-light ratio.  For
the massive, lower-redshift clusters, the
angular size of R$_{vir}$ was larger than the CFHT-MOS
Field-of-View (FOV) ($\sim$ 10$'$ $\times$ 10$'$)
and several MOS pointings were required.  For these clusters, either a north-south or
east-west strip through the cluster center was observed.  Column 8 of Table 1 lists the configuration
of the MOS pointings.  The first number in the column is the number of
east-west pointings and
the second is the number of north-south pointings.  A 1 $\times$ 1 pointing is the
cluster center.  Column 9 lists what percentage of the area of a
circle with radius R$_{200}$, the radius at which the mean density
of the cluster exceeds the critical density by a factor of 200 (see Paper II for updated values of
R$_{200}$ for the CNOC1 clusters using a $\Lambda$CDM cosmology), is covered by the optical imaging
data.  Column 3 of Table 1 lists the number of cluster members (see
Carlberg et al. 1996 for membership criteria) with spectroscopic redshifts
within R$_{200}$.  
\newline\indent
In order to overlap as much of the optical imaging and spectroscopic data as possible, we designed a pointing
strategy for the ONIS observations based on the locations of the
CFHT MOS fields.
To obtain images with similar coverage
as the $\sim$ 10$'$ $\times$ 10$'$ CFHT MOS (although only $\sim$ 8$'$ $\times$
8$'$ are useful for photometry and spectroscopy due to a significant radial distortion
and vignetting)
using the $\sim$ 2.5$'$ $\times$ 5$'$  FOV of ONIS we used a 3 $\times$ 2 
pointing pattern for each MOS field.  This gives us an ONIS mosaic with a FOV
of $\sim$ 7.5$'$ $\times$ 7.5$'$, which is similar to the usable
portion of the MOS.  Each pointing consists of
$\sim$ 4 dithers  of $\sim$
6$''$ to 10$''$, and each dither consists of 1-3 internal coadds with
exposure times of 25-45 seconds.  The nature of the mosaicking is such
that there is significant overlap of the pointings in the north-south direction and
a strip in the center of each field has a higher signal-to-noise
than the rest of the mosaic.  The depth of the images at the lower
signal-to-noise parts still reaches the limit of the spectroscopic
observations, so we make no correction for the slightly uneven depth
of the images.  The PISCES camera has a FOV of 8$'$ $\times$ 8$'$,
nearly the same size as the MOS, and therefore we made a single PISCES
pointing for each MOS pointing in MS0440+02.  The PISCES pointings
consist of 5 dithers of $\sim$ 5$''$ with a 40 second exposure time.

\section{Data Reduction}
Reduction of the ONIS data was done using the IRAF packages PHIIRS and
PHAT (see Hall, Green, \& Cohen, 1998).  Reduction of the PISCES data was done
with a modified version of the ipipe reduction package (Gilbank et al., 2003).  Both of these
packages employ the standard techniques for reducing NIR imaging, and here we summarize only the most important steps in the process.  
\newline\indent
The first step in the reduction is to make a first-pass dark subtraction, flat-fielding, and sky estimate 
for all frames.  Frames are then registered and coadded into a first-pass
mosaic of the cluster.  Object finding is performed on this mosaic, and from
this an ``object mask'' is made.    Flat-fielding and sky
subtraction are then redone using only pixels that do not appear in
the the object mask. This step ensures that faint galaxies not
detected in a single frame do not cause the sky level to be overestimated. 
\newline\indent
All ONIS images are flattened with dome flats (for the PISCES images we use the 
value of the sky itself to flat-field).  Whenever possible, sky subtraction
is done using a ``running sky''.  This procedure involves calculating
the sky from a fixed number of frames taken before and after the frame
for which the sky is being subtracted.  Determining the sky from a small number of frames 
improves the photometric accuracy of the data because the
moderate-duration sky variations (of the order 10 minutes in the NIR) can be removed accurately.  Unfortunately, using a small
number of frames to compute the sky also results in a larger
Root-Mean-Squared (RMS)
background in the final mosaic because shot noise is significant when using only a few frames.  
Our criteria for choosing the
best number of frames for sky subtraction was as follows:  The sky was
first computed using all the frames from a cluster (we termed this a ``quick sky'').
The quick sky was subtracted from all
frames, and the final mosaic made.  The RMS noise and standard
deviation of that image are then recorded.  We then repeat sky subtraction using the running sky technique.  The final sky we adopt is the one that uses the fewest
number of frames and gets an RMS background for the final image that is within 1$\sigma$ of the quick
sky image.  For most clusters this results in a sky computed from 
$\sim$ 16-32 frames (i.e., approximately 5 - 10 minutes before and
after each frame was observed).  Once the final sky has been
subtracted from each frame,  we adopt an airmass extinction term of
-0.08 mag for the K-band, and all frames are rescaled by 1.08 $\times$ $<$airmass$>$ of
the observation. As an
example of the data quality, we show a portion of the
final mosaics of one of the highest-redshift (MS0016+16) and
lowest-redshift (Abell 2390) clusters in the sample in Figures 1 \& 2.  
\subsection{ONIS Pattern Removal}
\indent
All frames obtained using the ONIS instrument have a lined pattern that appears in the bottom-half of the image.  This pattern is probably caused by extra noise
generated when the camera is read out.  The pattern is well-fit by a
variable-amplitude sinusoid curve
with a period 6 times the width of the chip.  We remove the pattern by fitting
each original frame (before flat-fielding and sky subtraction) with a sinusoid and then subtracting the fit from the image.  In all
cases the fits are good and all images are subsequently eye-checked
to make sure nothing which might significantly alter the photometry
is subtracted.  The
final mosaic for each cluster is made from the final sky-subtracted,
pattern-removed frames.
\subsection{PISCES Distortion Correction}
The PISCES camera has a radial distortion in the focal plane.  The
pattern is different each time the camera is mounted on the telescope
and therefore must be corrected using stars in the science images.  We
correct this pattern using code developed by McCarthy et al. (2001).
Comparing the positions of objects on the optical images with the
distortion-corrected PISCES images shows that this correction works
extremely well.  
\begin{figure*}
\epsscale{1.0}
\plotone{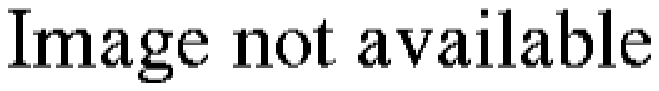}
\caption{\footnotesize K-band image of the central region of
  MS0016+16 ($z$ = 0.55), one of the highest-redshift clusters in the
  sample.  The image has a FOV of 3.5 $\times$ 3.0 arcmin, and seeing of
  1$\farcs$1.  North is to the top, east is to the left. }
\end{figure*}
\begin{figure*}
\epsscale{1.0}
\plotone{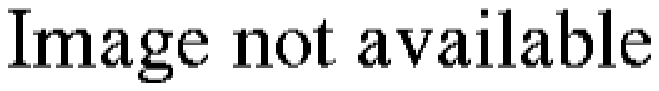}
\caption{\footnotesize K-band image of the central region of Abell
  2390 ($z$ = 0.22), one of the lowest-redshift clusters in the
  sample.  The image has a FOV of $\sim$ 2.5 $\times$ 3.5 arcmin, and seeing of
  1$\farcs$2.  North is to the left, east is to the bottom. }
\end{figure*}

\subsection{Object Finding and Photometry}
The $g$ and $r$-band optical images are significantly deeper than the K-band images for every
cluster.  Therefore we match the optical catalogue positions to the K-band
images using the IRAF task {\it xyxymatch}.  The significant 
pincushion of the CFHT-MOS is corrected using tabulated data supplied by CFHT.  In
most cases this allows for excellent matches of the 
whole optical catalogue.  In the few cases where a slight offset
remains, a higher order
correction is computed and then eye checked.  A final eye check is
made for objects that may have been so red as to be apparent only in
the K-band images.  Very few such objects were
found ($\sim$ 1-2 per MOS
field).  Those that are found are not faint Extremely Red
Objects (EROs); they tend to be objects undetected in the optical image because they are
obscured by the bleeding of charge from a saturated star.  No attempt is made
to account for these objects because there are so few.  Furthermore, these galaxies do
not have redshifts and therefore will have negligible impact on the results of the
study.
\newline\indent
Photometry and star-galaxy classification are done on all objects using
Picture Processing Package (PPP, Yee 1991).  PPP identifies objects
as either stars or galaxies based on analysis of their growth curve.
The {\it r}-band images are the deepest for all clusters and
therefore the object classification is based exclusively on the
$r$-band images.
\newline\indent
Galaxy colors are nominally computed using a 3$''$ diameter aperture magnitude.  
However, PPP determines whether this aperture is appropriate based on the
object's growth curve.  If the growth curve appears ``normal''
(i.e., it is monotonically increasing, but has a monotonically decreasing derivative)
then an actual 3$''$ aperture is used; however, if the
growth curve is abnormal (due to, e.g., crowding from another
object, or a cosmic ray hit) PPP uses the largest, non-contaminated
aperture to compute the color (we refer to this diameter as the optimal
aperture, d$_{opt}$).  Computing
colors using this technique is particularly useful for avoiding the crowding problems which can be
significant in the dense, central parts of galaxy clusters.
\newline\indent
Galaxy total magnitudes are determined by analyzing the shape of the
growth curve.  For faint small galaxies ($r$ $>$ 19) the magnitudes are extrapolated to a diameter of
8.5$''$ to account for seeing effects.  The extrapolation is done
using a reference Point Spread Function (PSF), determined from 
several bright stars in the field.  For brighter galaxies ($r$ $<$
19), variable apertures up to a maximum of 25.5$''$ are used.  The
size of the aperture is determined from the growth curve of the object.
This step is primarily employed to account for bright galaxies with large
angular diameters (i.e., foreground galaxies). 
A thorough discussion of photometry using PPP and a comparison with simulations
can be found in Yee (1991).  

\subsection{Photometric Calibration and Comparison with Previous
  Photometry}
\indent
Photometric standards from Persson et al. (1998) were observed
throughout all nights, at different
airmasses. There are not enough standards to solve for the
atmospheric extinction
coefficient; however,
comparing observations of the same standard at different air masses
shows that they are consistent with -0.08 value assumed for the
science frames.  Standards taken throughout the night were extremely
stable, so photometric zeropoints are determined for each night 
based on an average of all standards for that night.  The standard deviations in the
zeropoint calibration for the three ONIS nights are 0.022, 0.014, and
0.035 mag, and are determined from 3, 10, and 6 standard stars respectively.  The PISCES photometric zeropoint has a
standard deviation of 0.022 mag and is determined from 4 standard stars.  
\newline\indent
The cluster imaging is sufficiently wide that it contains enough bright
stars per cluster field to check the consistency of the photometry with the
2MASS point-source catalogue (Skrutskie et al. 2006).  The 2MASS photometric calibration is
excellent, having zeropoint variations that are less than 0.02 mag
across the entire survey.  For the purpose of determining any
differences in zeropoint, we compare the photometry of objects 13.5 $<$ K$_{s}$ $<$ 15.0
classified as stars by PPP in the cluster fields to the same objects in the
2MASS Point Source Catalogue.  We adopt a faint limit of K$_{s}$ $<$ 15.0
because the mean photometric uncertainty for an individual star at
K$_{s}$ $\sim$ 15 becomes fairly large in the 2MASS catalogue ($\sim$
0.1 mag).  We
choose a bright limit of K$_{s}$ $>$ 13.5 because stars begin to saturate at
this magnitude on the ONIS detector (the PISCES potential well is
slightly deeper and therefore a bright limit of K$_{s}$ = 
12.5 is adopted for MS0440+02).  The number of stars available for comparison in the fields of the
clusters varies from from 6 in MS1008-12 and MS1455+22 to 70 in Abell
2390.  In general, the agreement between our photometry and the 2MASS photometry is 
good. Eight of 15 clusters have median photometric differences of $<$ 0.05
mag, and 12 of 15 have differences of $<$ 0.1 mag.  The cluster with the
largest offset is MS0016+16, which is fainter than the 2MASS photometry
by 0.18 mag.  We noted the possibility of light cirrus in the logbook when
observing this cluster, and assume that this is the explanation for
the difference in zeropoint.  Table 2 lists the clusters, the number
of stars, and the mean difference
between our photometry and the 2MASS photometry.  Although the offsets
between our calibration and the 2MASS calibration are small, we have
chosen to recalibrate our photometry to the 2MASS photometry, rather than use the solutions
from the standard stars.  This provides consistent photometric zeropoint for the
entire cluster sample.  Once recalibration is
complete, the magnitudes of all galaxies are
corrected for Galactic reddening using the dust maps of Schlegel et al. (1998). 
\newline\indent
Three of the CNOC1 clusters (MS1358+62, MS0451-03, and MS1008-12) were also observed in the
study of Stanford et al. (1998, 2002; hereafter SED02).  As an
additional check on the consistency of our photometry,
we compare the magnitudes of all galaxies brighter than the 5$\sigma$
limiting depth of the shallowest observation (in all three cases our
data is shallower by $\sim$ 0.1 mag).  The top panels of Figures 3,
4, and 5 show plots which compare our total magnitudes with the SED02 total magnitudes.
The solid line in the plot has a slope of unity.  The
bottom panel of the plots shows the residuals.  Overall, the agreement
between our photometry and the SED02 photometry is poor.  The
median offset for MS1008-12, MS1358+62, and MS0451-03 is 0.089, 0.223,
and 0.305 mag respectively, with our photometry being fainter in all cases.
Objects which are extreme outliers ($>$ 1 mag) are likely to be bad matches, as we
use a relatively simple matching technique.
\newline\indent
The systematic differences in the photometry are much larger than the
zeropoint uncertainties ($\sim$ 0.02 mag for 2MASS, and $\sim$ 0.03
mag for the SED02 data) and they are also larger than the even the largest
difference between our own photometric calibration and the 2MASS
calibration (0.18 mag).  This makes it difficult to understand the
cause of the discrepancy.  The data reduction procedure 
employed by SED02 is nearly identical to our own, and therefore it is
unlikely that this causes any systematics between the datasets.  The
most notable difference between the SED02 datasets and our own is that the SED02 observations were
done with different cameras on different telescopes: OSIRIS on KPNO 1.3m for MS1008-12, IRIM
on KPNO 2.1m for MS1358+62, and IRIM on KPNO 4m for MS0451-03 whereas
ours were done with the same instrument/telescope configuration. 
In our program,
the observations for MS1008-12 and MS1358+62 were taken consecutively on February
3. If there were some
systematic change during the observations (e.g., cirrus) it most likely would 
have affected both of these clusters, yet 
the photometric calibration from the standard stars matches the 2MASS
calibration for both these clusters very well ($<$$\Delta m$$>$ =
-0.003 and -0.039, for MS1008-12 and MS1368+62, respectively).
\newline\indent
It is interesting to note that the photometry for objects brighter
than K $\sim$ 15.0 agrees somewhat better than for objects fainter than K
$\sim$ 15.  Given that our zeropoints are determined using stars brighter
than this, it suggests that the SED02 photometry could be as consistent
with the 2MASS photometry as our own, and that the relative offset is
not a zeropointing
problem, but may be a systematic due to the different way magnitudes are measured.   
The cameras used by SED02 have a much larger pixel scale than ONIS
(0\farcs95/pixel, 1\farcs09/pixel, 0\farcs65/pixel for OSIRIS,
IRIM-2.1m, and IRIM-4m, respectively,
compared to 0\farcs288/pixel for ONIS).  Such a large pixel size means
that the some of their observations may be significantly undersampled if the
seeing was close to 1\farcs0 or better.
This could pose a problem because the
SED02 magnitudes were measured using the FOCAS (Valdes, 1982) code,
which determines magnitudes using isophotes.  SED02 then correct the
FOCAS magnitudes because FOCAS tends to produce total magnitudes which
are too faint.  It is possible that determining isophotal magnitudes
using undersampled data may result in a significant
scatter, because the location of the isophote is difficult to
determine.  We cannot be certain that this is the source of the discrepancy
over such a small area; however, because we use
primarily a single instrument, and our photometry agrees well with
2MASS photometry, we are confident of the photometric
accuracy of our own data. 
\begin{figure}
\epsscale{1.1}
\plotone{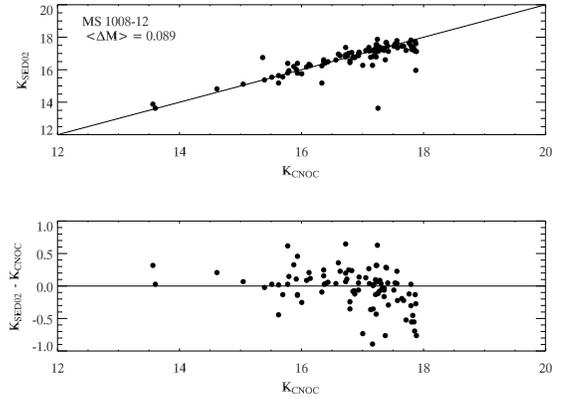}
\caption{\footnotesize Top Panel: Total magnitudes from the SED02
  photometry vs. total magnitudes from this photometry for the cluster
  MS1008-12.  The solid
  line has a slope of unity and intercept of zero.  Bottom Panel:
  Residuals from the top panel. }
\end{figure}
\begin{figure}
\epsscale{1.1}
\plotone{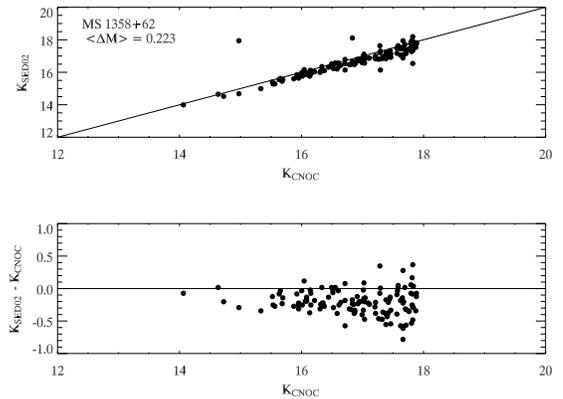}
\caption{\footnotesize As Figure 3 for the cluster MS1358+62. Objects that are
  significant outliers may be poorly matched. }
\end{figure}
\begin{figure}
\epsscale{1.1}
\plotone{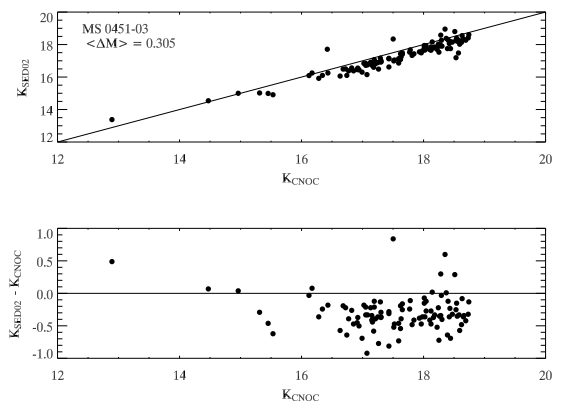}
\caption{\footnotesize As Figure 3 for the cluster MS0451-03.  Objects that are
  significant outliers may be poorly matched. }
\end{figure}
\section{Spectroscopic Selection Function}
The large number of spectroscopic redshifts available from the CNOC1
project allows us to measure the
K-band LF and density profiles with more confidence than by
statistical background subtraction; however, the cluster redshift catalogues are
sparsely sampled and a weighting scheme is required to correct for the spectroscopic selection function.  Galaxies
may preferentially have a measured redshift depending on their magnitude, color,
position in the cluster, or {\it z}; and the galaxy weights are a function of these
four parameters.
\newline\indent 
The weighting scheme that is employed is based on the one derived by the
CNOC1 and CNOC2 collaborations and is discussed thoroughly in the
respective papers (Yee
et al. 1996a, Yee et al. 2000).  The basic philosophy behind the weighting
scheme is that the distribution of galaxies with redshifts is representative of the
ones without redshifts in terms of the primary selection biases.  
The magnitude selection function (i.e., a brighter
galaxy is more likely to have a spectroscopic redshift) is
overwhelmingly dominant over the other 3 possible selection effects,
which can be considered ``secondary effects'' (Yee
et al. 1996a).
Incorporating the full set of weights in our computations has little
effect on the final results and for simplicity of interpretation we
ignore the secondary biases and use only the magnitude weights when
computing the cluster LFs.
\newline\indent
The spectroscopic weights are computed as follows.  Galaxies
with redshifts are compared to the total number of galaxies in running
bins of $\pm$ 0.25 mag for galaxies fainter than K$^{*}$ + 1.5, and in
running bins of $\pm$ 0.50 mag for galaxies brighter than K$^{*}$
+ 1.5.  K$^{*}$ for each cluster is estimated using a passive evolution
model (\S 6.3).  The weight for a galaxy is
then the inverse ratio of galaxies with redshifts to the total number
of galaxies in its
magnitude bin.  The weights for the
brightest cluster galaxies (BCGs) are sometimes not equal to 1, even
though the spectroscopy for the BCGs
 is 100\% complete.  This occurs because there are bright field galaxies
within the cluster field.  In order to avoid overestimating the
contribution from bright galaxies, the
spectroscopic weight of the BCG in each cluster is set equal to 1.  
\newline\indent
All galaxies within the cluster field are used to compute the weights
when we measure the cluster density profiles.  For the LFs, only galaxies within
R$_{200}$ are used to determine the weights.   We adopt this approach
for the LFs because the K-band imaging does not have the same coverage
as the 
optical imaging/spectroscopy in all clusters.  A few clusters have
K-band data to only R $\sim$ R$_{200}$ whereas others have coverage well
beyond R$_{200}$ and therefore have a larger proportion of field
galaxies with redshifts.  
Throughout the analysis in this paper, the determination of
cluster membership is done using the cluster redshift-space bounds calculated by
Carlberg et al. (1996).
\newline\indent
One potentially serious problem with the spectroscopic catalogue is that
it is {\it r}-band selected, yet it is being used to determine the
abundance of K-band selected galaxies.  If a cluster or field contains a
significant number of EROs which are redder than the cluster red-sequence
then they will be missing from the spectroscopic sample,
and could artificially inflate the K-band spectroscopic weights.
Although we
already verified qualitatively that there is not a significant number of EROs
in the cluster field (\S 3.3), one way to further confirm
there is no bias in the $r$-band selected spectroscopic catalogue is to compare the weights computed for
the {\it r}-band data to the K-band weights.  Figures 6 and 7 show plots
of the {\it r}-band weights and K-band weights for the clusters
MS1358+62 and MS0302+16.  MS1358+62 has the best spectroscopic completeness of
the sample, while MS0302+16 has the worst.  The weight functions show
the characteristic dip at bright magnitudes ($r <$ 20, K $<$ 16) and
the gradual fall-off at fainter magnitudes.  The dip at bright
magnitudes occurs because most of the bright galaxies are close
together in the cluster core and
getting slits on all of them is difficult, even with the multiple-mask
strategy of CNOC1.  The weight functions are
similar between the K-band and $r$-band for both clusters, and this
behavior is similar for all clusters in the sample.  Therefore, we
conclude that the $r$-selected spectroscopy is still
representative of the K-band sample of galaxies.
\newline  
\begin{figure}
\epsscale{1.1}
\plotone{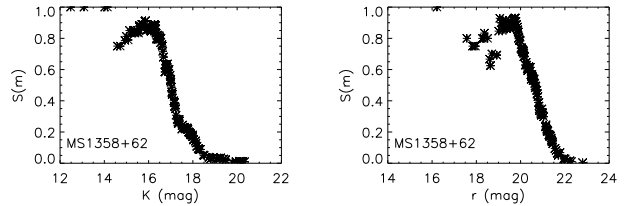}
\caption{\footnotesize An example of the inverse of the magnitude weights (S(m)) for the
galaxies in MS1358+62.  Spectroscopic coverage of this cluster is
very complete and the distribution of K-band weights is
similar to the $r$-band weights.}
\end{figure}
\begin{figure}
\epsscale{1.1}
\plotone{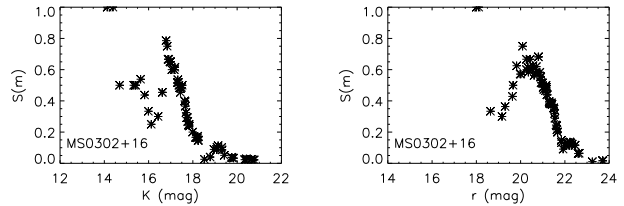}
\caption{\footnotesize As Figure 6 for the cluster MS0302+16.
  Spectroscopic coverage of MS0302+16 is
somewhat poor, however the distribution of K-band weights is still
similar to the $r$-band weights.
}
\end{figure}
\section{Cluster Density Profiles}
The mass density profiles and $r$-band number density profiles
for the CNOC1 clusters have already been measured using the 
spectroscopic dataset by several authors (Carlberg et al. 1997a,
1997b; van der Marel et al. 2000).  
Here we compute the K-band selected luminosity and number density profiles as a
comparison to check whether K-band selected galaxies (which better trace
stellar mass) produce profiles that are different from
$r$-selected galaxies.  Computing the luminosity and number density
profiles also allows us to compare the K-band density profiles of the higher-redshift CNOC1 clusters
to the K-band density profiles of lower-redshift clusters
(e.g., L04, R04). 
\newline\indent
The density profiles are constructed by stacking all 15 clusters into an
ensemble cluster.  Before adding galaxies to the ensemble cluster
the radius of each cluster is normalized by its R$_{200}$.
The number density profile is calculated by totaling the number of
cluster galaxies with K $<$ K$^{*}$ + 1 in circular shells, scaling each galaxy
by its spectroscopic weight ($\S$4).  The luminosity profile is
computed in the same manner using the K-corrected, and evolution corrected
(see $\S$6.1) luminosity of galaxies with K $<$ K$^{*}$ + 1.  
\newline\indent
The clusters do not have homogeneous radial coverage and most are
observed in a strip running through the cluster center.
Therefore, the total counts or total luminosity in each
shell is multiplied by the relative coverage of the shell before
converting to a surface density.  Some clusters also have
coverage that extends to a larger R than others.  To reduce the
noise at large R from poorly sampled clusters, the
contribution from individual clusters is truncated at the radius where the actual coverage
of the shell is less than 10\% and each data point in the ensemble
cluster is weighted
by the number of clusters that contribute to it.  The data for the ensemble
cluster extends to 1.5R$_{200}$.
\newline\indent
In the top panel of Figure 8 we plot the cluster number density
profile (including the BCGs) and in the bottom panel we plot the
luminosity density profile (excluding the BCGs).  The data are fit to a projected
NFW (Navarro et al. 1997) profile of the form derived by Lokas \&
Mamon (2001)
\begin{equation}
\Sigma(R) =  Ac^2 g(c)\frac{ 1 - |x^2 - 1|^{-1/2} C^{-1}(1/x) }{ (x^2 - 1)^2 },
\end{equation}
where c is the concentration parameter, x $\equiv$ R/r$_{s}$ and
\begin{equation}
g(c) = [ \ln(1+c) - c/(1+c) ]^{-1},
\end{equation}
\begin{equation}
C^{-1}(x) = \left\{ \begin{array}{l} 
\mbox{cos}^{-1}(x) \qquad \mbox{if R $>$ r$_{s}$}\\
\mbox{cosh}^{-1}(x)\qquad \mbox{if R $<$ r$_{s}$}.
\end{array} \right.
\end{equation}
\indent The parameters A and c are fit using the Levenberg-Marquardt least-squares technique for
$\chi^2$-minimization (Press et al. 1992).
The best fit concentration parameter for the galaxy number density
profile is c$_{g}$ = 4.13 $\pm$ 0.57.  This agrees well with the
c$_{g}$ = 3.7 calculated by Carlberg et al. (1997b) using the $r$-band
photometry.  It is also similar to the best-fit concentration of the K-band
luminosity profile, c$_{l}$ = 4.23 $\pm$ 0.70.  
\newline\indent
Interestingly, a good fit for the luminosity profile 
can only be obtained when the luminosity of the BCGs
is excluded.  The asterisk in the bottom panel of Figure 8 shows the value of the
central point in the luminosity density with the BCGs included.  This central part is better described by a
power-law fit (dashed line) with an index of n = -1.38 $\pm$ 0.03.
The power-law can describe the luminosity profile with
BCGs included
approximately as well as the NFW profile does with the BCGs removed (reduced-$\chi^2$ = 2.97 and
2.79 respectively).  It is perhaps surprising that the {\it number} density
profile can be described by an NFW profile when the BCGs are included,
but the {\it luminosity} density profile can only be fit when the BCGs are
removed.  This illustrates the unique nature of the BCG luminosity in
the context of the formation of large-scale structure.  The
distribution of K-band light closely follows the distribution of
K-band-selected halos
throughout the cluster which suggests the average luminosity per halo is
roughly constant for cluster galaxies, except the central galaxy which
is by far the brightest galaxy residing in a single halo.  This 
is also intuitively supported by the cluster LFs ($\S$6 \& $\S$7)
where the luminosity distribution of cluster galaxies is well-fit by
a Schechter function, except for the BCGs which are much brighter than
the rest of the population and are more abundant than a Schechter
function would predict.
\newline\indent
How do the luminosity and number density profiles compare to the cluster mass
profiles?  van der Marel et al. (2000) computed the mass profile of the CNOC1 clusters
with detailed Jeans equation analysis.  They showed that a variety of
generalized density profiles fit these data well, with the best-fitting NFW model having 
c$_{DM}$ = 4.17 (unfortunately, no error-bar is quoted for this value).  This is nearly identical to the
concentration of the K-band number and luminosity density profiles and
suggests that both the
stellar mass contained in galaxies (excluding the BCGs) and the
stellar-mass selected subhalo population, tightly trace the
dark matter mass.  The same conclusion was made by both
Carlberg et al. (1997b) and van der Marel et al. (2000) using the $r$-band selected number
density.   However, if the BCGs are included in the luminosity
profile, it appears that stellar mass may dominate over dark matter in
the cluster core.  The same trend is seen by Sand et al. (2004) using
a sample of 6 clusters in the same redshift range (0.17 $< z <$ 0.44)
selected for having radial arcs.
Sand et al. (2004) use a combination of strong-lensing, BCG velocity
distribution and BCG luminosity to model the ratio of
luminous and dark matter in cluster cores.  Unfortunately, the
K-band photometry and the van der Marel et al. (2000) velocity
dispersion profile do not reach the resolution of the Sand et al.
(2004) sample at small radii and therefore we can only
remark that the CNOC1 profiles appear to be consistent with their result.
\newline\indent
A comparison of the cluster number density profiles to those measured at lower
redshift shows that the K-band light in the somewhat more massive, higher-redshift, CNOC1 clusters is more
concentrated, at the 2$\sigma$ level.  L04 find c$_{g}$ =
2.88$^{+0.21}_{-0.10}$ for an ensemble of 93 $z <$ 0.1
clusters using K-band photometry from 2MASS.   Similarly, R04 measure
$<$c$_{g}$$>$ = 2.83 $\pm$ 0.56 in the K-band for the 9 CAIRNS clusters.
Unfortunately, because these samples and the CNOC1 clusters have
different masses and are at different redshifts, understanding why their concentrations 
differ is not entirely straightforward. 
\newline\indent
N-body simulations show that c (for dark matter)
is: 1) higher in lower-mass halos; 2) higher in halos which collapse
first; and 3) for virialized halos becomes higher with decreasing redshift as the halo accretes
mass (e.g., Navarro et al. 1997; Wechsler et al. 2002).
However, the simulations show that the dependence of
c on 1) and 3) is relatively weak, especially for cluster-mass halos. 
Given that the concentration parameter is most strongly dependent on the collapse
epoch of the halo, the factor of $\sim$ 1.5
difference between the concentrations of these samples suggests an earlier
collapse time for the CNOC1 clusters.   Furthermore, the fact that
the CNOC1 clusters are more massive and observed at an earlier epoch than the L04 and CAIRNS
clusters would seem to support this interpretation.
Nonetheless, these simulations are for dark matter
only, and do not measure the concentration of stellar light.
Therefore, it remains possible that 
the difference in concentrations between the samples could be
caused by a redshift evolution in the concentration of
stellar light, rather than different collapse epochs for the two samples.
\newline\indent
Interestingly, the concentration of galaxy number density and dark matter are
the same in the CNOC1 clusters (within the precision of our data), yet 
this is not seen in the local clusters.
Both L04 and R04 find that the K-band selected c$_{g}$ is
smaller than c$_{DM}$ for their clusters.  Their results
also agree with the simulations of Nagai \& Kravtsov
(2005) who find c$_{DM}$ $>$ c$_{g}$ for a set of
8 clusters in a full hydrodynamical simulation.  Five of the simulated
clusters have c$_{DM}$ which is a factor of $\sim$ 2 larger
than c$_{g}$, while 2 clusters have c$_{DM}$ $\approx$ c$_{g}$.  The
8th cluster has c$_{g}$ $>$ c$_{DM}$.   These ratios are similar to
the CAIRNS clusters, where the mean ratio between
c$_{DM}$ and c$_{g}$ (for the NFW model) is 2.3 $\pm$ 0.67.  While the errors in
c$_{g}$ and c$_{l}$ for the CNOC1 clusters are somewhat large, they still exclude the
possibility that the number density of K-band-selected galaxies is less concentrated than the matter by
a factor of 2 at the 3$\sigma$ level (c$_{g}$/c$_{DM}$ = 0.99 $\pm$
0.14 and c$_{l}$/c$_{DM}$ = 1.01 $\pm$ 0.17), suggesting a real difference
between the moderate and low-redshift cluster samples.
\newline\indent
If we assume that the ratio of c$_{DM}$/c$_{g}$ $\approx$ 2 at $z =$ 0
seen by R04 and L04 is universal for galaxy clusters, then this
suggests the possibility of an evolution in the relative c$_{g}$
and c$_{DM}$ with redshift.  
For c$_{DM}$ to increase faster than c$_{l}$ or c$_{g}$
would require a preferential radial segregation of dark matter and
light during the accretion process.   One consideration is that such a
segregation could be mimicked by a redshift
evolution in the radial distribution of
cluster galaxy stellar populations.  While possible, this is unlikely to be the case for our profiles which are computed using K-band
light, which is fairly insensitive to star-formation properties or
dust.  Furthermore, they are computed using a
cut of K $<$ K$^{*}$ + 1.0 where K$^{*}$ is very well determined and
consistent with simple passive evolution ($\S$6.3).  
\newline\indent
The simulations of Nagai \& Kravtsov (2005) show that accreted galaxies near
the cluster core have had $\sim$ 70\% of their total halo mass
stripped since being accreted by the cluster, whereas those near the
virial radius have lost only $\sim$ 30\%.  Since stellar mass tends to
be tightly bound within a dark matter halo, the process of tidal
stripping of dark matter subhalos
could plausibly cause a differential evolution in c$_{DM}$
and c$_{g}$ for galaxy clusters.  The tidally stripped dark matter
of the infalling galaxies may sink to the center of the cluster halo
and increase c$_{DM}$, while the stellar mass will remain
bound to the subhalo and continue to orbit within the cluster halo, where it
preferentially spends more time at the perimeter.
\newline\indent 
While this is a possibility, we note that with the current dataset it is impossible to untangle whether the
difference in the relative concentrations of dark matter and stellar
mass between the moderate and low-redshift samples
is caused by such an evolution or simply because the ratio of c$_{DM}$ to c$_{g}$ is not universal for all
cluster masses at all epochs.
It would be interesting to compare these results with simulations that
trace the redshift evolution of the cluster c$_{DM}$ and c$_{g}$.
It would also be useful to compare to
simulations that have the same c$_{DM}$ as the CNOC1 clusters. 
We note that all of the Nagai \& Kravstov (2005) clusters have higher
concentrations than the CNOC1 clusters and that the two
clusters with the lowest concentrations (which are similar to
the CNOC1 clusters) have a ratio of c$_{DM}$/c$_{g}$ $\approx$ 1.  
\begin{figure}[htbp]
\epsscale{1.10}
\plotone{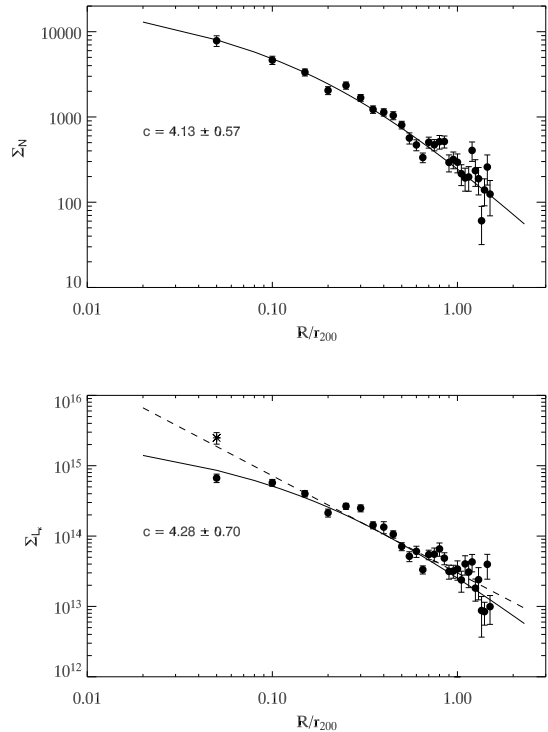}
\caption{\footnotesize Top Panel: Surface number density per virial area of K-band
  selected galaxies vs. radius. The solid line is the best-fit NFW
  profile. Bottom Panel: Surface K-band luminosity density per viral
  area vs. radius.  The asterisk is the luminosity density calculated when the BCGs
  are included.  The solid line is the best-fit NFW profile excluding
   the BCGs.  The dashed line is the best-fit power-law ($\gamma$ =
  -1.38 $\pm$ 0.03) including the BCGs.
   }
\end{figure}
\section{Cluster Luminosity Functions}
In this section we compute the K-band LFs for cluster
galaxies.  Our technique for creating stacked, k-corrected LFs
using clusters at different redshifts and correcting for imaging of varying depths
is discussed in \S 6.1.  The method outlined there is used to create all LFs in
the remainder of the paper ($\S$6 and $\S$7). 
In \S 6.2 we construct an ensemble cluster LF using
all 15 clusters centered at $z \sim$ 0.3.  This allows us to fit the
faint-end slope of cluster LF ($\alpha$)
as well as compare it to field LFs at the same redshift.  In
\S 6.3 and \S 6.4 we examine the redshift evolution of K$^{*}$ as well as its
dependence on cluster mass using a fixed value of $\alpha$.  
\subsection{Technique for Constructing Luminosity Functions}
LFs are constructed by stacking the sample of 15 clusters
into either a single cluster (\S 6.2, \S 7.2) or else into 3 redshift (\S
6.3, \S 7.3) or
mass (\S 6.4, \S 7.4) bins with 5 clusters per bin.  Many clusters have enough
spectroscopy to determine their individual LFs reasonably
well; however, stacking the clusters improves the statistical errors.  
The LFs are constructed by counting the number of
cluster galaxies (multiplied by their spectroscopic
magnitude weights) within
R$_{200}$ in bins of 0.25 mag.  Most clusters have some galaxies that lie
outside the projected R$_{200}$ but have velocities consistent
with cluster membership; these galaxies are excluded from the
LFs.  R04
showed that the K-band cluster LF becomes fainter by $\sim$ 
0.5 magnitudes from the virialized to infall regions in local clusters.
There is insufficient coverage of the infall region to compute a
separate LF and therefore the 
LFs are constructed using only galaxies within the virialized region. 
\newline\indent
The cluster BCGs are not removed when computing the luminosity
functions.  The inclusion the BCGs
results in inflated reduced-$\chi^2$'s for the fits (because
the BCGs do not follow a Schechter function); however, similar
to the optical cluster LFs of de Propris et al. (2003), we find that removing
the BCGs has no significant effect on the fitted value of K$^{*}$ ($\sim$
0.01 mag).  Therefore, the BCGs are included as an indication of
their abundance relative to other cluster galaxies.  In the
determination of all
LFs (\S 6 and \S 7) no attempt is made to account
for the incomplete coverage of R$_{200}$ in some clusters.  Most
clusters have full coverage of R$_{200}$, and those that do not have a
strip of observations across the cluster which provides reasonably
equal sampling of the cluster center and the periphery.  
\newline\indent
The clusters are at different redshifts and therefore stacking them
requires that the magnitudes of galaxies within each cluster be ``redshifted'' to
a common redshift.  For the ensemble LFs
(\S 6.2, \S 7.2) and the LFs for clusters of different
masses (\S 6.4, \S 7.4) we use absolute magnitudes at a common redshift of $z$ = 0.296, the median redshift of
the sample.  This requires the
computation of a distance modulus, k-correction and differential
evolution correction for each galaxy.  When we compute the redshift
evolution of the LFs (\S 6.3, \S 7.3) we keep the
conventions of the literature and use apparent magnitudes.  This requires only a differential
distance modulus and differential k-correction.
\subsubsection{k-Corrections}
The k-corrections are taken from the models of Poggianti
(1997) who list corrections for E, E1 (an early-type with longer duration
episode of star-formation, which we consider an S0), Sa, and Sc types.  The K-band
k-corrections are fairly independent of spectral type; 
however, there are small differences, and therefore determining a
spectral type for each galaxy is preferable.  The availability of the optical
photometry allows us to compute {\it g} - $r$ colors and estimate the spectral
type of each galaxy using a simple model for how the colors of different
spectral types evolve with redshift.   We have fit the
{\it g} - $r$ vs. $r$ color-magnitude relation of each cluster (Paper III)
using the biweight estimator (Beers et al. 1990).  For the purpose of
spectral classification of the galaxies, we assume that all cluster galaxies that are $<$ 0.1
magnitudes bluer than the cluster red-sequence are early-type galaxies.
The redshift/color models of Fukugita, Shimasaku, \& Ichikawa (1995)
are then used as template
colors for the remaining galaxies.  Rather than directly using the colors in
their models, the color of an early-type galaxy at the
appropriate redshift is determined empirically from the cluster color-magnitude relation, and then the
{\it differential} colors from the models are used to determine whether
galaxies are S0, Sa, or Sc types.  This approach minimizes any
systematics which might be caused by incorrect
normalization of the
models.  Furthermore, because the colors between the 4 spectral types are
significantly different, it implies an error of $\pm$1 spectral
type at most.  This is fairly inconsequential as the
k-corrections differ by only 0.02 - 0.05 mag across all spectral-types
for galaxies at 0.2 $< z <$ 0.55. 
\subsubsection{Evolution Correction}
Previous studies have shown that there is strong luminosity evolution with
redshift in the K-band for both field (Drory et al. 2003, Feulner et
al. 2003, Pozzetti et al. 2003) and cluster (dP99) galaxies.
Therefore,  a differential evolution correction must be included in
the LFs to ``evolve'' 
higher and lower redshift clusters to the appropriate redshift.  We
use the evolution corrections listed in Poggianti (1997).  The Poggianti models consist of 15 Gyr old galaxies in
a $q_{0}$ = 0.225, H$_{0}$ = 50 km s$^{-1}$ cosmology.  We have
adjusted the evolution corrections by assuming the
galaxies are 13.7 Gyr old at the present, and then mapped the
$q_{0}$ = 0.225 cosmology on to our own by comparing lookback times.  Once this
correction is made, the Poggianti evolution corrections are in excellent
agreement with our own measurement of the luminosity evolution for
cluster galaxies (see \S 6.3).  Computing the LF
at the median redshift of the clusters, rather than correcting to $z$
= 0 allows smaller evolution 
corrections (which are of order 0.08 to 0.18 mag depending on
spectral-type and cluster redshift) and therefore, the choice of
evolution model does not strongly affect the results from the LFs.
\subsubsection{Completeness}
Not all clusters have photometry which is complete
to the same absolute magnitude; hence, to maximize the depth of the
stacked LFs we adopt the approach of Schechter (1976).
The clusters are ranked by limiting absolute
magnitude, and the limiting magnitude of the stacked LFs is set by the
depth of the deepest cluster.  Clusters are then added to the stacked
LF in the order of deepest to shallowest.  The counts at magnitudes fainter than
the completeness limit for a shallower cluster are extrapolated
from the stacked LF of the deeper clusters.
Using this technique means that the faintest bins in the stacked
LF are scaled versions of the faint-end 
of the deepest clusters, whereas the bright end is determined from all
clusters.  While not strictly correct, it maximizes the
information on the bright-end of the LF, where the statistics are
poorest.  Furthermore, most of the clusters
are complete to approximately the same depth ($\sim$ K$_{*}$ + 2)
with only the two highest redshift clusters (MS0016+16 and MS0451-03)
being notably shallower ($\sim$ K$_{*}$ + 1); hence, only the
faintest bins are affected by this approach.  
Because we do not use statistical background subtraction
we assume the errors in each
bin of the LF to simply be Poisson errors.  The errors are computed  
{\it before} the faint-end of the LF is scaled.  
\newline\indent
The fitting of all LFs is done using
the Levenberg-Marquardt
algorithm for $\chi^2$ minimization (Press et al. 1992).  Errors for the parameters are estimated from the covariance
matrix.  
\subsection{Ensemble Luminosity Function and Comparison of the Cluster
  and Field Luminosity Function at Moderate Redshift}
Here we construct a composite LF from all 15 CNOC1
clusters.  By stacking the clusters we reduce the statistical errors
and are able to make a good measurement
of both K$^{*}$ and $\alpha$ using imaging of only moderate depth.
\newline\indent 
Figure 9 shows the composite LF for all clusters
centered at $z$ = 0.296, the median redshift of the sample.  The best-fit Schechter function parameters are K$^{*}$ = -24.53 $\pm$ 0.15
and $\alpha$ = -0.84 $\pm$ 0.08.  The
fit parameters for this LF as well as all other LFs computed in this
paper are listed in Table 3.  If we compare this LF
to the local K-band cluster LF 
measured by L04 we find the following.
1) The faint-end of the LF
has not evolved from $z$ = 0, to $z$ = 0.3.  L04 find that the
best-fit $\alpha$ for their LF is $\alpha$ = -0.84 $\pm$ 0.02,
which is identical to the best-fit for the CNOC1 clusters.
2) The evolution in K$^{*}$ is
consistent with a passive luminosity evolution of the stellar population.  L04 measure K$^{*}$ =
-24.02 $\pm$ 0.02 for their clusters, which implies K$^{*}$ for the CNOC1 clusters
is 0.51 $\pm$ 0.15 mag brighter at $z$ = 0.3.  The  Poggianti evolution
model (adapted to our cosmology, \S 6.1) predicts 0.31 magnitudes of
luminosity evolution from $z$ = 0 to $z$ = 0.3 for an early-type
galaxy, which is smaller, but consistent
with the observed evolution.  A more
detailed investigation and discussion of the redshift
evolution of K$^{*}$ is presented in \S 6.3.
\newline\indent
The cluster LF shows no
significant change (other than the
passive aging of the stellar populations) with redshift between $z$ =
0 and $z$ = 0.3; however, it is worthwhile to consider
whether it depends on environment at $z$ = 0.3.  In the
local universe the K-band LFs of the field
and cluster environments are different.   B01, Lin et al. (2003), L04,
R04, Ramella et al. (2004), and
Kochanek et al. (2003) have all shown that the field and cluster have
similar faint-end slopes, but that  K$^{*}$ is brighter (by $\sim$ 0.2 - 0.4
mag) in clusters.  
Recently, several $z >$ 0.1 field K-band LFs have become available in the
literature, and we can now compare the cluster and field LFs at $z =$ 0.3.  
\newline\indent
Pozzetti
et al. (2003) use the K20 survey with a set of $\sim$ 500
spectroscopic redshifts to determine the evolution of the K-band
field LF from $z$ = 0.2 to $z$ = 1.5.  They find
that the evolution of K$^{*}$ to $z \sim$ 1 is
consistent with a luminosity evolution of $\Delta$K$^{*}$ = -0.54 $\pm$
0.12.  The MUNICS survey group combined their
K-band photometry with spectroscopic (Feulner et al. 2003) 
and photometric (Drory et al. 2003) redshifts to determine the
evolution of the K-band field LF from z = 0 to $z$ $\sim$ 1.
Feulner et al. (2003) find that the K-band field galaxy LF evolves by
$\Delta$K$^{*}$ = -0.70 $\pm$ 0.30
magnitudes from z = 0 to $z \sim$ 1, while Drory et al. (2003) find similar
($\Delta$K$^{*}$ $\sim$ -0.5 to -0.7 mag) results.  
\newline\indent
The faint-end slopes of the field LFs are steeper than the $\alpha$ =
-0.84 $\pm$ 0.08
measured for the CNOC1 clusters.  In their $z$ = 0.2 - 0.65 redshift bin, Pozzetti et
al. (2003) find that $\alpha$ = -1.25$^{+0.25}_{-0.20}$, and Feulner et al. (2003) assume that
$\alpha$ = -1.1 in their $z$ = 0.1 - 0.3 and $z$ = 0.3 - 0.6 redshift bins.  
Unfortunately, the correlation between K$^{*}$ and $\alpha$ makes the 
comparison of K$^{*}$ from LFs that use different
values of $\alpha$ difficult to interpret.  
Therefore, we refit the
cluster LF forcing $\alpha$ to be fixed at -1.1.
This fit is shown as the dashed line in Figure 9.
With $\alpha$ fixed at -1.1, the best-fit value for the cluster LF is
K$^{*}$ = -24.93 $\pm$ 0.04 (the smaller error bar arises because
$\alpha$ is held fixed).
\newline\indent
The field LFs have coarse redshift bins, and therefore
we can only compare the cluster and field LFs as an
average over a broad redshift range.
For example, the lowest redshift bin in the Pozzetti et al. (2003) study is 0.2 $< z <$
0.65, which spans the entire redshift range of the CNOC1 clusters.
The mean redshift of the CNOC1 clusters is similar to the
mean redshift of the two lowest bins from Feulner et al. (2003), so we
average those values to K$^{*}$ = -24.68 $\pm$ 0.26 at $z$ $\sim$
0.3.  Comparing this to our value shows that K$^{*}$ is 0.25 $\pm$ 0.26 magnitudes brighter in clusters than the
field at $z$ $\sim$ 0.3.  If we compare our LF to the
$z$ = 0.2 - 0.6 redshift bin of Pozzetti et al. (2003) who obtain K$^{*}$
= -24.87$^{+0.54}_{-0.77}$ we find that K$^{*}$ in clusters is 0.05$^{+0.54}_{-0.77}$
magnitudes brighter than in the field.  Both these values agree
within the large error bars;
however, the error bar on the Feulner et al. (2003) value is
significantly smaller, and consistent with the results from the much
larger photometric redshift study of Drory et
al. (2003).  Therefore, we adopt $\Delta$K$^{*}$ = -0.25 $\pm$ 0.26
between field and cluster at $z \sim$ 0.3 as our best result.
\newline\indent
Unfortunately, the errors bars on the field values of K$^{*}$ are
quite large, and clearly any difference between the cluster and field
at moderate redshift is of the order of the error bars or less.
Despite the large error bars, it is
interesting to note that the difference in K$^{*}$ between cluster and
field galaxies at $z$ = 0.3 is similar to that at $z \sim$ 0.  This
suggests that the processes which cause the K-band luminosity
function, and by corollary the stellar mass function, to be brighter
in clusters (e.g., hierarchical growth of cluster galaxies from
mergers) probably have
already occurred by $z \sim$ 0.3, and that there is
little differential evolution since then.  A larger field K-band study,
with finer redshift bins and better constraints on K$^{*}$ would be useful for investigating
this further.
\begin{figure}
\epsscale{1.1}
\plotone{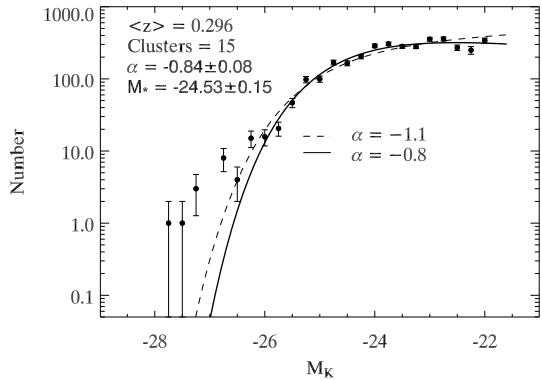}
\caption{\footnotesize Stacked LF for all 15
  clusters, corrected to a redshift of $z$ = 0.296.  The solid line is
  the best-fit Schechter function (K$^{*}$ =
  -24.53 $\pm$ 0.15, $\alpha$ = -0.84 $\pm$ 0.08).  The dashed line is the best-fit Schechter
  function with a fixed $\alpha$ = -1.1 (K$^{*}$ = -24.93 $\pm$
  0.04).  The bright end of the LF diverges from a Schechter
  function because the BCGs are included.}
\end{figure}
\subsection{Redshift Evolution of the Cluster Luminosity Function}
In \S 6.2 we showed that the composite LF for all 15
CNOC1 clusters, when compared with $z \sim$ 0 clusters, is consistent with no evolution in the faint-end slope
and passive evolution of the stellar populations.  Here we make a more
detailed examination of the redshift
evolution of K$^{*}$ by separating the clusters into 3 redshift bins with 5 clusters per
bin.  The LFs are computed at the mean redshift of the 5
clusters within each bin.  This corresponds to redshift bins of $z$ = 0.21,
0.29, and 0.46.  For these LFs we use apparent rather than absolute magnitudes so that we can make a direct comparison to the
LFs of dP99.   
\newline\indent
When the LFs are separated into redshift bins, they do not have
enough depth to obtain meaningful constraints on both K$^{*}$ and
$\alpha$ (especially in the highest redshift bin).  Therefore, we hold $\alpha$ constant and fit 
only K$^{*}$ and $\phi^{*}$. Rather than assume $\alpha$ = -0.84, the
best-fit value for the ensemble LF, we adopt the
value of $\alpha$ = -0.9 assumed by dP99, which is consistent with
our best-fit. 
Because of the strong correlation between K$^*$ and $\alpha$,
assuming the same faint-end slope also allows for a straightforward
comparison of K$^*$'s from different studies.
We note that given that there is a $\sim$ 10\% uncertainty in $\alpha$ from the
combined LF ($\S$6.2), the error on K$^*$ for a given $\alpha$
is an underestimate of the total error budget.
To obtain an estimate of the total error budget for K$^{*}$ which 
includes the uncertainty in $\alpha$, we refit the LFs with values of
$\alpha$ that range from -0.76 to -0.92 (i.e., enclosing  the 1$\sigma$
error bar in $\alpha$ from $\S$6.2).  This refitting results in values of
K$^{*}$ that are +0.15 magnitudes fainter when $\alpha$ = 0.76 
and -0.05 magnitudes brighter when $\alpha$ = 0.92.  These additional
deviations are comparable to the fitting errors with fixed $\alpha$s (see Table 3). 
\newline\indent
In Figure 10 we plot the LFs and the Schechter function fits for the 3 redshift bins.  
The redshift evolution of K$^{*}$ for the CNOC1 clusters is compared
with the dP99 values in Figure 11.  The lines in the figure are
models of single-burst populations with a $z_f$ = 1.0, 1.5, 2.0, 2.8, and 5.0
constructed using the Bruzual \& Charlot (2003) code.  The model is a
0.1 Gyr duration single-burst with solar metallicity followed by an
exponentially declining star-formation rate with $\tau$ = 0.1 Gyr.  The model is
normalized to K$^{*}$ = -24.02 at $z$ = 0, the L04 value which was measured 
using a faint-end slope ($\alpha$ = -0.84) that is similar to the
$\alpha$ = -0.9 assumed for our clusters and the dP99 clusters.   Figure 11
demonstrates that the stellar populations in CNOC1 clusters are consistent with a passively
evolving population formed at z $>$ 1.5; however, the values of K$^{*}$ for the CNOC1 clusters are
significantly fainter than those of the dP99 clusters. 
If we compare the $z$ = 0.21, 0.29, and 0.46 bins of the CNOC1
clusters to the $z$ = 0.20, 0.32, and 0.46 bins of dP99, K$^{*}$ for the
CNOC1 clusters is fainter by 0.36, 0.19, and 0.76 mag, respectively, and are 0.44
mag fainter on average.  
\newline\indent
We have considered several possible explanations for the very different values
of K$^{*}$ between the studies, the most likely of which is the
systematic differences in the photometry.  In \S 3.4 we noted that for the 3 clusters that overlap
our sample and the SED02 sample (the data used to compute the dP99
LFs), our photometry is systematically
fainter by 0.21 mag.  If the entire datasets differ by this much, then
this accounts for approximately half of the discrepancy in K$^{*}$.  
\newline\indent
It is possible that the different radial coverage in the two samples
is partially responsible for the difference in K$^{*}$.  R04 showed that the K-band cluster
LF becomes fainter from the virial to infall region.  If a such
radial dependence also exists within the virial region, then this may partially explain
the discrepancy because the dP99 data cover only the central region of
the clusters ($\sim$ 0.5 - 1 Mpc) whereas our 
observations cover out to R$_{200}$ ($\sim$ 1 - 2 Mpc) for most clusters.  We test this
possibility by 
recomputing the LFs using
only galaxies within 0.5 Mpc of the cluster center.  K$^{*}$ from those luminosity
functions is brighter; however, by only $\sim$ 0.1 magnitudes,
which is smaller than the discrepancy, even if photometry accounts for
a portion of it.  We also consider the possibility that the size of the magnitude bins used for the
LFs  may influence the value of K$^{*}$.  dP99 use larger 
bins (0.5 mag) which might bias the value of K$^{*}$ to brighter
values because of the poor statistics at the
bright-end of the LF.  When we recompute our LFs using
0.5 mag bins instead of 0.25 mag bins we find that this has no
effect on K$^{*}$.  
\newline\indent
Perhaps the most significant difference in the methods used to derive the luminosity
functions is that we use spectroscopic
redshifts, whereas dP99 use statistical background
subtraction.  In principle, both methods work equally well, however,
the statistical method requires the stacking of a large number of
clusters, because cosmic variance in the background can cause large
errors in the LFs.  Two of the three dP99 redshift bins that compare
with ours have only a few clusters in each bin (3, 9, and 2
respectively).  A higher than average background in the cluster field might result in an
overestimate of the number of cluster galaxies.  
We cannot be conclusive as to why the value of K$^{*}$ in the CNOC1
clusters is significantly fainter than in the dP99 sample; however,
our simple comparisons suggest that the size of the magnitude bins and the different radial coverage
between the samples has little effect on K$^{*}$.  Most
likely, the difference is caused by differences in the
photometry (\S 3.4), and possibly because of the different techniques
used for background subtraction.
\newline\indent
Comparing the LFs to the K$^{*}$ = -24.02, $\alpha$ = -0.84 LF of L04
shows there is an evolution of $\Delta$K$^{*}$ = -0.35
$\pm$ 0.06 mag from $z$ = 0 to $z$ = 0.46.  This agrees well with the
passive evolution predicted from the Bruzual \& Charlot single burst,
$z_{f}$ = 2.8 model ($\Delta$K$^{*}$ = -0.40) and the passive
evolution from the Poggianti (1997) model ($\Delta$K$^{*}$ = -0.39).  We conclude that the CNOC1
clusters agree well with the scenario where the bulk of the
stars in galaxies are formed at high-redshift and evolve passively thereafter.  Furthermore,
the close relation between the K-band light and
stellar mass of a galaxy suggest that 
the stellar mass function of K $<$ K$^{*}$ + 2 cluster galaxies is
unchanged up to $z = $ 0.3.  
\newline\indent
It is difficult to understand how no evolution in $\alpha$ (\S 6.2)
and purely passive evolution of K$^{*}$ is compatible with the L04 HOD data which suggest a significant
number of mergers in this redshift range.  Even if mergers populate
all parts of the LF appropriately as to maintain the
overall shape, K$^{*}$ would have to be fainter than
passive evolution at higher-redshift to account for the breakup of
galaxies into their progenitors.  It is possible that  
the reduction in luminosity of galaxies at high redshift due to
breakup could be offset by increased amounts of
star-formation which correspondingly brighten the galaxy, and
therefore mimic passive evolution; however, such a 
scenario seems contrived, and the most reasonable interpretation of the data
is that galaxies in massive, relaxed, X-ray-selected clusters do not
experience a significant number of mergers between 0 $< z <$ 0.3.  This may
not be a surprising result, as the high velocity dispersion of galaxies in the cluster
environment makes merging difficult.  The passive evolution of the
LFs at moderate redshift does not rule out the
possibility that mergers play a role in cluster galaxy evolution;
however, it suggests that if they are important, they most likely
occur in higher-redshift systems that are in the process of relaxing (e.g., MS1054+03, Tran et al. 2005), rather than
massive virialized clusters at $z \sim$ 0.3.
\begin{figure}
\epsscale{1.1}
\plotone{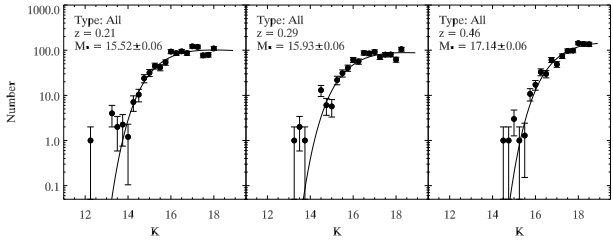}
\caption{\footnotesize Stacked LF for clusters in
  three different redshift bins.  Each LF is composed
  from a stack of 5 clusters that have been corrected to the mean
  redshift listed in each panel.  The solid-line is the best-fit Schechter
  function assuming $\alpha$ = -0.9.  The bright-end of the LFs
  sometimes diverge from a Schecheter
  function because the BCGs are included. }
\end{figure}
\begin{figure}
\epsscale{1.2}
\plotone{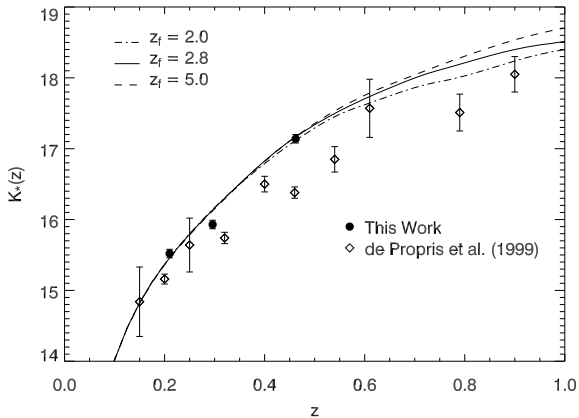}
\caption{\footnotesize Evolution of K$^{*}$ as a function of
  redshift.  Solid points are the CNOC1 cluster LFs
  (Figure 10) and the open diamonds are clusters from dP99.  The 
  long-dashed, triple-dot-dashed, dot-dashed, solid, and dashed lines are
  single burst models with $z_{f}$ = 1.0, 1.5, 2.0, 2.8, 5.0 respectively.  The
  models have been normalized to the low-redshift value of K$^{*}$
  from the L04 study (see text).  The bright galaxies in the CNOC1
  clusters are consistent with a scenario where the bulk of the
  stellar mass is formed at high
  redshift and evolves passively thereafter.  }
\end{figure}

\subsection{Luminosity Functions of Different Mass Clusters}
\indent
Here we separate the CNOC1 sample into 3 mass bins using cluster
dynamical masses determined from the velocity dispersions of Carlberg et
al. (1997a, see Paper II for values of M$_{200}$ computed using a
$\Lambda$CDM cosmology) and investigate the dependence of the K-band LF on
cluster mass.    
The mean masses of the three bins are $<$M$_{200}$$>$ = 2.73 $\times$ 10$^{14}$
M$_{\odot}$, 5.96 $\times$ 10$^{14}$ M$_{\odot}$, and 1.34 $\times$ 10$^{15}$  
M$_{\odot}$, and we designate these low, mid, and high-mass bins, respectively.
\newline\indent
Figure 12 shows the LFs for the 3 mass bins, as well as the
best Schechter function fits, again using a fixed $\alpha$ = -0.9.
The values of K$^{*}$ for the three mass bins are K$^{*}$ =
-24.51 $\pm$ 0.08, -24.59 $\pm$ 0.06, -24.52 $\pm$ 0.05 for the low,
mid and high mass bins respectively.  This shows that there is no significant
dependence of K$^{*}$ on cluster mass over the 
one order-of-magnitude range in mass covered by the CNOC1 clusters.
\newline\indent
Interestingly,  L04 do not find the same result in local clusters.
They divide their sample of 93 clusters into high and low-mass
subgroups (using masses determined from X-ray temperatures) and 
find that $\alpha$ is similar between the two groups
($\alpha$ = -0.84 $\pm$ 0.03 and
-0.81 $\pm$ 0.04 for high-mass and low-mass clusters, respectively); however, K$^{*}$ is brighter
by 0.16 $\pm$ 0.07 mag in high-mass clusters
(K$^{*}$ = -24.10$\pm$ 0.04 in high-mass clusters vs. K$^{*}$ = -23.94
$\pm$ 0.06 in low-mass clusters).  The L04 high-mass clusters have a
mean M$_{200}$ similar to the mid-mass CNOC1 clusters, while
their low-mass clusters have a mean M$_{200}$ similar to the low-mass
CNOC1 clusters, and therefore the corresponding difference between those mass bins in
the CNOC1 sample is $\Delta$K$^{*}$ = -0.08 $\pm$ 0.10 mag.   
\newline\indent
The dP99 clusters have a similar redshift range as the CNOC1 clusters,
and similarly, they do not show a dependence of K$^{*}$ on cluster mass.  
Although masses for their clusters were unavailable at the time, dP99
divided their sample into high and 
low optical richness and high and low X-ray luminosity (L$_{x}$) subgroups.  Both
optical richness and L$_{x}$ are correlated with cluster mass
(Yee \& Ellingson 2003; although there is some scatter)
and therefore these subgroups can be considered roughly as 
high and low-mass subgroups.  Similar to our result, dP99 find that within the errors, K$^{*}$ is the
same between high and low richness and high and low L$_{x}$ clusters from $z$ =
0.15 to $z$ $\sim$ 1.0.  
\newline\indent
Given the uncertainty of our measurement of the difference in K$^{*}$
between different mass bins, our result is 
consistent at $<$ 1 $\sigma$ with both no dependence of K$^{*}$ on
cluster mass (dP99) or a very weak ($\sim$ 0.1 mag) dependence of K$^{*}$
on cluster mass (L04).   
\begin{figure}
\epsscale{1.1}
\plotone{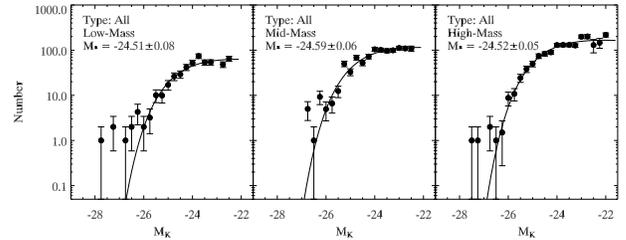}
\caption{\footnotesize Stacked LF for clusters in
  three different mass bins.  Each LF is composed
  from a stack of 5 clusters.  The solid-line is the best-fit Schechter
  function assuming $\alpha$ = -0.9. The bright-end of the LFs
  sometimes diverge from a Schechter
  function because the BCGs are included.  There appears to be no
  correlation between K$^{*}$ and mass for over the mass range covered
  by the CNOC1 clusters.}
\end{figure}
\section{Luminosity Functions of Different Spectral-Types}
\indent
The analysis in \S 6 demonstrated that the K-band cluster
LF shows no evolution in $\alpha$ and only passive
evolution in K$^{*}$ from $z = 0$ to $z = 0.3$.  Because of the close correlation
between K-band light and stellar mass it also suggests
no strong evolution in the stellar mass function of cluster galaxies over
this redshift range.  However, studies of the evolution of
morphology-density relation (Postman et al. 2005, Smith et al., 2005,
Dressler et al., 1997), as well as the star-formation rate in clusters
(e.g., Balogh et al. 1999, Poggianti et al. 1999), and the
Butcher-Oemler effect (Ellingson et al. 2001) suggest that there is
significant evolution in the morphological and star-formation
properties of the cluster galaxy population over the same
redshift range.  Therefore, it seems likely that that the LFs of the early and
late-type populations will evolve differently over this
redshift range, even though the combined
LF of cluster galaxies shows only passive evolution.
\newline\indent
This possibility can be addressed directly within the CNOC1 sample
using the spectroscopy.  Ellingson et al. (2001)
performed Principal Component Analysis (PCA) on the spectroscopic
data and classified galaxies into three broad spectral-types.  In this section we use the
PCA analysis and examine the
K-band LF of these spectral-types.  In \S 7.1 we briefly
summarize the PCA decomposition.  In \S 7.2 we construct a composite
LF for the spectral
types using all 15 clusters so that we can fit $\alpha$.  Using the best-fit values of
$\alpha$ we study the redshift evolution of K$^{*}$ and its
dependence on cluster mass for the different spectral types in \S 7.3
and \S 7.4.
\subsection{PCA Decomposition}
\indent
Here we present a brief discussion of the PCA analysis.   
A thorough explanation of the fitting method
and reliability of the spectral-typing is presented in Ellingson et
al. (2001).
The principal components of a galaxy's spectrum are determined by
decomposing it using three galaxy types as eigenvectors: Elliptical,
Emission-line, and Balmer-line.  The template spectra used for these
types are composite
spectra drawn from the Las Campanas Redshift survey (Shectman et al. 1996). 
In the PCA analysis, each eigenvector is assigned an amplitude from 0 to 1, based on how well it
represents the spectrum being fit.  The total amplitude of all three
eigenvectors adds up to 1.  While this analysis is
simplistic, it is quite effective at providing a reasonable
quantitative measurement of the principal component of a galaxy's spectrum.
Using the amplitudes of the PCA decomposition we divide the galaxies
into 2 broad spectral-types:
Early (ELL), and Emission + Balmer (EM+BAL).  Galaxies with ELL $>$ 0.5
are considered ELL, and those with ELL $<$ 0.5
are considered part of the EM+BAL class.  This provides
a simple way to identify star-forming or recently 
star-forming galaxies from those which are dominated by
absorption-lines and are likely to have been quiescent for at least a few
Gyr.  It is important to note that this analysis is a spectral
analysis, not a morphological one.  Galaxies which have early-type
morphologies may still be considered EM+BAL galaxies if they show the
appropriate spectral features (in fact, the BCGs in the highest
redshift clusters all show emission lines and therefore do not fall
into the ELL category).  Separating galaxies by spectral-type
(rather than morphology) is
similar to the analysis done by B01 for low-redshift clusters
and groups (although they use line indices, not PCA), and therefore
allows easy comparison between moderate and low-redshift clusters. 
\subsection{Ensemble Spectrally-Types Luminosity Functions}
In Figure 13 we show the LFs for ELL and EL+BAL
classes.  Immediately obvious is the difference between the
faint-end slopes of the two LFs.  The best-fit
faint-end slope for the EM+BAL galaxies is $\alpha$ = -0.95 $\pm$ 0.27,
while for the ELL galaxies it is $\alpha$ = 0.17 $\pm$ 0.18.  
This difference indicates significant redshift evolution in the
faint populations of these spectral-types because in the local field and local clusters 
their faint-end slopes are nearly identical.  
\newline\indent
In the local field, Kochanek et al. (2001) showed that for morphologically-typed galaxies, early and late-types
have faint-end slopes of -0.92 $\pm$ 0.10 and -0.87 $\pm$ 0.09
respectively.  Bell et al. (2003) performed the same analysis
using spectral-types from
Sloan Digital Sky Survey (SDSS) spectroscopy and 2MASS photometry and
found that $\alpha$ was similar across the types, but that it was
slightly shallower for the early-types.  In local clusters B01 examined a set of
galaxies in the field, group, and cluster environment using 2MASS
photometry and spectroscopic redshifts from the Las Campanas Redshift
Survey.  They split their sample into emission-line (EL) and no-emission-line (NEL) types
and although the error bars are large, they find that local cluster EL and NEL galaxies have comparable
faint-end slopes ($\alpha$ = -1.18 $\pm$ 0.76 for EL and $\alpha$ =
-1.28 $\pm$ 0.50 for NEL).  They also find that these values of $\alpha$ are
similar to their local field and group values for EL and NEL galaxies. 
\newline\indent
Our result suggests a strong decrease in the faint ELL population in clusters
from $z = 0$ to $z = 0.3$, whereas the faint EM+BAL population
remains mostly unchanged.  We can make a rough estimate of the relative decrease of
K$^{*}$ $<$ K $<$ K$^{*}$ + 2 ELL galaxies between $z =$ 0 and $z = 0.3$
by integrating the LFs.  Assuming the B01, $z = 0$ value of $\alpha$ in
local clusters ($\alpha$ = -1.18 $\pm$ 0.76) and our own best-fit value for $z =$
0.3 clusters ($\alpha$ = 0.17 $\pm$ 0.18) and integrating the number
of galaxies between K$^{*}$ $<$ K $<$ K$^{*}$ + 2 for these different
values of $\alpha$, we find that the number of ELL galaxies with
K$^{*}$ $<$ K $<$ K$^{*}$ + 2  decreases by a factor of 3.8 over this
redshift range.  Considering the large error bar in the B01 value of
$\alpha$, we also compare to an $\alpha$ = -0.92 LF (the Kochanek et
al. 2001 value for field early-types).  If the local cluster
early-type population has the same faint-end slope as the field, then the relative decrease in
K$^{*}$ $<$ K $<$ K$^{*}$ + 2 galaxies is a factor of 2.6. 
\newline\indent
One potential concern with our ELL LF is that it may suffer from
selection effects caused by difficulty obtaining successful redshifts
for faint, absorption-line systems.  Extensive tests on the
completeness of the spectroscopy (Yee et al. 1996) and PCA analysis
(Ellingson et al. 2001) show that the spectroscopy is complete to
the adopted limits; however, given the implications it is worth
exploring this result further.  As a check that the measured decrease
in $\alpha$ is not caused by selection effects in the faintest bins, we refit the ELL LF
using galaxies more than a full magnitude brighter than the
spectroscopic completeness limit.  This LF has a slightly steeper
slope and a larger error bar ($\alpha$ = -0.04 $\pm$ 0.24); however,
it is $>$ 3$\sigma$ different than the EM+BAL limit.  Furthermore, within the
error, it is completely consistent with the measurement using the full
spectroscopic catalogue.  From this we conclude that a
significant difference in the number of faint ELL vs. EM+BAL galaxies
does exist, and is not a selection effect.
\newline\indent
A decreasing number of faint,
non-starforming galaxies with increasing redshift is expected in the
``downsizing'' scenario of galaxy formation (Cowie el al. 1996).
Moreover, the same decrease is predicted by studies of
the fundamental plane in the local universe.  Nelan et al. (2005)
show that the typical age of the stellar populations of
low-luminosity early-types is $\sim$ 4 Gyr.  If these galaxies form in
the monolithic collapse scenario, then this suggests that they would
be star-forming galaxies at z $>$ 0.3 and would not populate
the faint-end of the ELL LF.  
\newline\indent
The same trend has also been observed as a decrease in the number of faint red-sequence galaxies at high
redshift.  Kodama et al. (2004), De Lucia et al. (2004), and Tanaka et
al. (2005) all show that
clusters at $z > 0.7 $ have fewer faint red-sequence galaxies than their
low-redshift counterparts. 
\newline\indent
Interestingly, the decrease in faint ELL galaxies in the CNOC1 clusters is
not met with a corresponding increase in the number of EM+BAL galaxies.  The
faint-end slope measured from these galaxies is in good agreement with
the local field and cluster values.  This suggests that if the
faint-end of the ELL LF is built-up between $z = 0.3$
and $z = 0$ from the quenching of star-formation in faint EM+BAL galaxies,
that these galaxies must be replenished in order to maintain the faint-end
slope.  This could naturally be explained by a scenario where faint,
star-forming galaxies are continuously accreted from the field and
gradually transformed into quiescent galaxies.
\begin{figure}
\epsscale{1.1}
\plotone{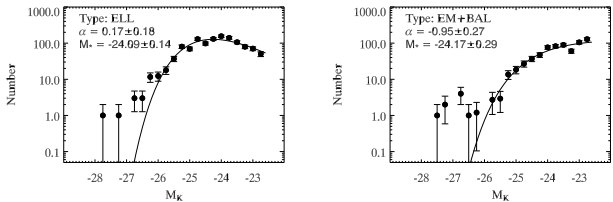}
\caption{\footnotesize Left Panel:  Ensemble LF
  for galaxies in all 15 clusters with spectra classified as ELL.  The
  LF is corrected to $z$ = 0.296.
  Right Panel:   Same as the left panel for galaxies classified as
  EM+BAL.  The bright-end of the LFs diverge from a Schechter
  function because the BCGs are included.
}
\end{figure}

\subsection{Redshift Evolution of Spectrally-Typed Luminosity Functions}
\indent
It is also useful to test whether the value of K$^{*}$ from the
spectrally-typed LFs shows any differential evolution
with redshift.  Here we compute LFs for the spectral-types at different redshifts in the identical manner
as $\S$6.3.  For these LFs we hold the faint-end
fixed using the values measured in \S 7.2
($\alpha$ = 0.2 for ELL galaxies and $\alpha$ = -0.9 for EM+BAL
galaxies).  Figure 14 shows the resulting LFs, and the
Schechter function fits.  Surprisingly, the ELL LFs have 
fainter values of K$^{*}$ than the EM+BAL galaxies.  The same result
is not seen in local K-band cluster and field studies. B01 show that in
local clusters the NEL galaxies are
0.60 mag brighter than the EL galaxies and Kochanek
et al. (2001) show that they are 0.55 mag brighter in
the local field.
While K$^{*}$ is brighter for the EM+BAL galaxies in the CNOC1
clusters, this does not mean that the average EM+BAL galaxy is
brighter than the average ELL because K$^{*}$ and $\alpha$ are correlated.
Shallower values of $\alpha$ typically result in fainter values of
K$^{*}$.  As an example, if we refit the ELL galaxies using a much steeper faint-end
slope $\alpha$ = -0.5 (the value measured for the stellar mass function of
early-type galaxies in the local universe by Bell et al. 2003), then we find that K$_{*}$ is
brighter for the ELL galaxies by 0.51,
0.50, and 0.59 mag for the $z$ = 0.21, 0.29, and 0.46 bins, and that,
similar to the local K-band studies, 
the ELL galaxies have values of K$_{*}$ which are brighter
than the EM+BAL galaxies.
\newline\indent
We can compare the evolution of K$^{*}$ with some simple models of
galaxy evolution.  The left and right panels of Figure 15 shows a plot of K$^{*}$ vs. $z$ for the
ELL and EM+BAL types, respectively.  The solid-line in the right panel is a $z_{f}$ = 2.8 single-burst
model normalized to the B01 value of K$^{*}$ = -23.31 for EL
galaxies at $z$ = 0.  The dashed line in the right panel is a stellar
population constructed with the Bruzual \& Charlot (2003) code which
forms half its stars in a single-burst at $z$ = 2.8, 
and the other half with a constant star-formation rate of 5M$_{\odot}$
yr$^{-1}$.
The solid line in the left panel shows the $z_{f}$ = 2.8 single-burst
model for the ELL galaxies.  Unfortunately, the ELL passive 
evolution model can not be normalized to the K$^{*}$ computed 
for the NEL LF by B01 because it is measured using $\alpha$ =
-1.28 and this value is very different from the $\alpha$ = 0.2
that we use.  Instead, the ELL single-burst model is
normalized to pass through the $z$ = 0.29 value of K$^{*}$ for ELL
galaxies. 
\newline\indent
While it is difficult to make robust conclusions from Figure 15, it is
worth noting that both the ELL and EM+BAL types are consistent with
single-burst, passive evolution models.  
This result, combined with the fact that the total cluster K$^{*}$
evolves passively would be consistent with a scenario where 
the bulk of the stellar mass in bright cluster galaxies is formed at
high-redshift and the dominant evolution thereafter is the passive
aging of the stellar populations, {\it regardless of
  spectral-type}.  It also suggests there is no inconsistency between
studies which find a significant change in
the morphology, color and star-formation properties of the cluster galaxy population at $z$ $>$ 0.1, 
and studies which have shown that the evolution of the their stellar
population is primarily passive (e.g., Stanford et al. 1998, van Dokkum
et al. 1998).  
Even though the LFs of the
ELL and EM+BAL galaxies at $z = 0.3$ change significantly by
$z = 0$, there is no corresponding change the total cluster LF. This suggests that the
transformations in morphology and
color/spectral-type which occur to cluster galaxies over the same redshift range are ``superficial'' -
they have little effect on the overall stellar
mass of the galaxies which transform. 
\begin{figure}
\epsscale{1.2}
\plotone{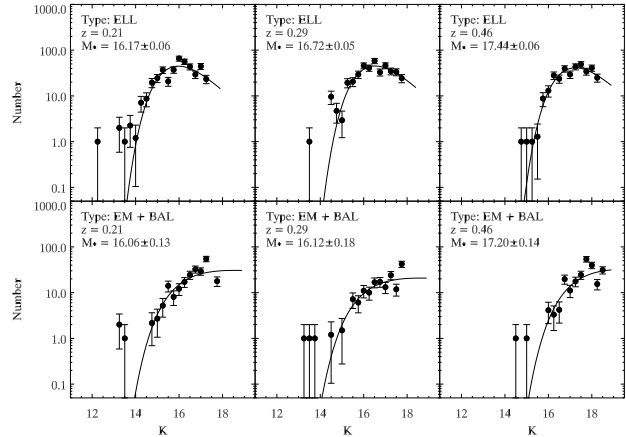}
\caption{\footnotesize Top Row: LFs at increasing
  redshift for galaxies classified as ELL.  Bottom Row: Luminosity
  functions at increasing redshift for galaxies classified as EM+BAL.
  Each LF is constructed from a stack of 5 clusters.}
\end{figure}
\begin{figure}
\epsscale{1.1}
\plotone{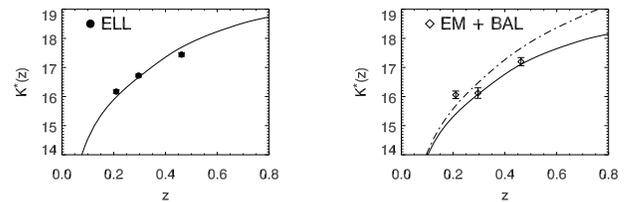}
\caption{\footnotesize Left Panel: Redshift evolution in K$^{*}$ for ELL
  galaxies.  The solid line is passive evolution model normalized to
  pass through the $z$ = 0.29 point.  Right Panel: Redshift evolution in K$^{*}$ for EM+BAL galaxies.  The
  solid-line is a passive evolution model normalized to the B01 EL
  K$^{*}$.  The dot-dashed line is a single-burst + constant star-formation
  rate model with the same normalization.}
\end{figure}

\subsection{Dependence of Spectrally-Typed Luminosity Functions on Cluster Mass}
In \S 6.4 we showed that there was no dependence of the overall K-band
LF on cluster mass for the mass range covered by the
CNOC1 clusters.  Here we test whether the LFs of galaxies of different spectral-types varies in clusters of different mass.
\newline\indent
Figure 16 shows the LFs for the ELL and EM+BAL types in the 3 mass bins used in
$\S$6.4.  The LFs are computed by again assuming $\alpha$ =
0.2 for ELL galaxies and $\alpha$ = -0.9 for EM+BAL galaxies.  Figure
16 demonstrates that there is no significant change in K$^{*}$ for
both spectral types
across all three mass bins, with the possible exception of a slight
trend (at $\sim$ 2$\sigma$ level) with mass for the ELL galaxies.  This suggests that any properties of the
cluster environment that depend on the cluster mass (e.g., ram-pressure stripping
efficiency, tidal forces) do not drastically alter the LF of different spectral types.  It might be expected that there
would be fewer faint, EM+BAL galaxies than bright EM+BAL galaxies in higher-mass clusters,
as they would be more susceptible to having their star-formation
truncated by tidal forces or ram-pressure stripping.  However, to the
depth of our LFs, we see no such reduction in the number of these
galaxies.
\newline\indent
Also, given that K$^{*}$ becomes brighter through  
the field, group, and cluster environment in the local universe (Kochanek
et al. 2001, B01, Ramella et al. 2004, L04, R04) it is surprising
that there is no significant difference for the spectral-types across
the cluster mass spectrum.  
\begin{figure}
\includegraphics[width=85mm]{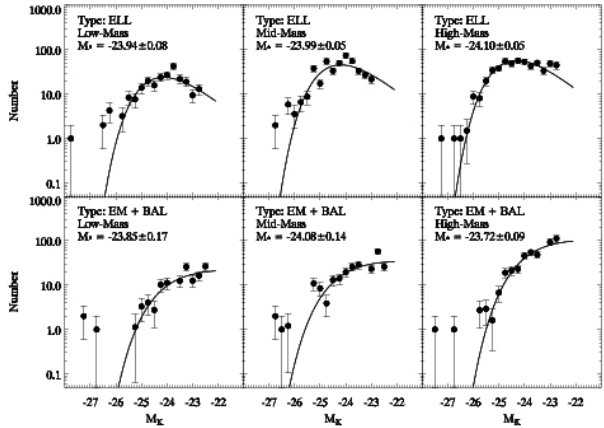}
\caption{\footnotesize Top Row: LFs for galaxies
  classified as ELL in clusters of different mass.  Bottom Row: Luminosity
  functions for galaxies classified as EM+BAL in the same clusters.
  Each LF is constructed from a stack of 5 clusters,
  and corrected to a mean redshift of $z$ = 0.296}
\end{figure}

\section{Summary}
\indent
We have presented K-band photometry for 15 moderate-redshift CNOC1 clusters
with extensive optical spectroscopy.
Our results show that both the luminosity and number density profiles
of the clusters are well-fit by NFW profiles with concentration
parameters of c $\sim$ 4.  Furthermore, comparison with the dynamical
mass analysis for the same clusters shows that for massive,
moderate-redshift, X-ray selected clusters, K-band light closely traces the dark
matter mass at R $<$ 1.5R$_{200}$, except possibly in the cluster core.  The galaxy number and luminosity
densities of the CNOC1 clusters are more concentrated than
local clusters, and this is likely caused by an earlier
collapse epoch for the higher-mass, higher-redshift CNOC1 clusters.
Furthermore, the ratio of c$_{DM}$ to c$_{g}$ for the CNOC1 clusters is
less than in local clusters and we suggest that this is the product of
either a non-universal ratio of c$_{DM}$ to c$_{g}$ for clusters, or else
a relative evolution in c$_{DM}$ and c$_{g}$ with redshift. 
\newline\indent
Analysis of the cluster LFs show that the evolution of K$^{*}$ between 0.2 $< z <$ 0.5
is consistent with a scenario where the majority of the stellar
mass in cluster galaxies forms at high redshift ($z_{f} >$ 1.5) and evolves
passively thereafter.  The faint-end slope of the LF shows
no evolution from the value measured in local clusters.  These results
imply that the stellar mass
buildup of individual galaxies through major mergers is negligible in
massive, X-ray selected clusters from $z$
= 0.3 to $z$ = 0.  
\newline\indent
We have also compared the K-band luminosity
functions at moderate redshift in different environments ranging from
the field to high-mass clusters.  Our results suggest that K$^{*}$ may
increase slightly in brightness from the field to the 
cluster environment at moderate-redshift; however, the error bars on
the field K-band LFs are large and therefore, the
data could also be consistent with no change.  Unlike local clusters, it
appears that for the CNOC1 clusters there is no
correlation between K$^{*}$ and the dynamical mass of the cluster.
\newline\indent
By dividing galaxies into star-forming/recently starforming (EM+BAL)
and non-starforming (ELL) types we examined the individual LFs
of these types.  The faint-end slope of the ELL LF
is significantly shallower than the faint-end slope for EM+BAL
LF.  Comparing the value of $\alpha$ for ELL galaxies
to local field and
cluster LFs suggest that the number of passive K$^{*}$ $<$ K
$<$ K$^{*}$ + 2 galaxies in clusters decreases by a factor of $\sim$ 3
from $z = $ 0 to $z$ = 0.3.  These results are consistent with
``downsizing'' in the cluster galaxy population.
\newline\indent
The spectrally-typed LFs also show that K$^{*}$ for both
ELL and EM+BAL galaxies is consistent with a passive evolution
scenario and this, in tandem with the passive evolution of the
combined cluster LF, suggests that the bulk of the stellar mass in both
types of galaxies is formed at high-redshift, and that subsequent
star-formation or changes in morphology do not affect the overall stellar mass of the
galaxies.  The spectrally-typed LFs also show that
K$^{*}$  for both EM+BAL and ELL classes is independent of the mass of the cluster that they
reside in.  
\acknowledgements
A.M. would like to cite useful conversations and help from David Gilbank and
Kris Blindert which significantly improved the quality of the data
analysis.  A. M. acknowledges financial support from the National
Science and Engineering Research Council (NSERC) in the form of PGSA and PGSD2
scholarships.  The research of H.K.C.Y. is supported by grants from
Canada Research Chair Program, NSERC and the University of Toronto.
E.E. acknowledges support for this research from the National Science Foundation under
Grants No. 9617145 and 0206154.  P.H. acknowledges support from NSERC.

\begin{deluxetable}{lccccccrr}
\tablecolumns{8}
\tablecaption{Summary of Observational Data for the CNOC1 Cluster Sample}
\tablewidth{6.3in}
\tablehead{\colhead{Cluster} & \colhead{{\it z}} &
\colhead{N$_{spec}$} & \colhead{K$_{lim}$} &  \colhead{Seeing}
&
\colhead{$g$/$r$ area}  & \colhead{K area} 
 & \colhead{MOS} & \colhead{\% R$_{200}$}  \\ 
\colhead{} & \colhead{} &
\colhead{(R $\leq$ R$_{200}$)} & \colhead{(mag)} &   \colhead{($\prime\prime$)}
&
\colhead{($\Box^\prime$)}  & \colhead{($\Box^\prime$)} 
 & \colhead{Fields} & \colhead{Obs}   \\ 
\colhead{(1)}& \colhead{(2)}& \colhead{(3)}& \colhead{(4)}&
\colhead{(5)}& \colhead{(6)}& \colhead{(7)}& \colhead{(8)}& \colhead{(9)}
}
\startdata
A2390       &  0.2279 & 140 & 18.08 & 1.2 & 269.56 & 249.01 & 5 $\times$ 1 & 60.9 \\  
MS0016+16   &  0.5466 &  52 & 18.57 & 1.1 &57.58 &  57.58 & 1 $\times$ 1 & 100.0 \\
MS0302+16   &  0.4246 &  26 & 19.12 & 0.9 &63.27 &  62.62 & 1 $\times$ 1 &  100.0 \\
MS0440+02   &  0.1965 &  33 & 19.01 & 1.4 &182.02 & 168.46 & 3 $\times$ 1 & 89.0\\
MS0451+02   &  0.2010 &  76 & 18.66 & 0.9 &253.41 & 173.75 & 4 $\times$ 1 & 53.6  \\
MS0451-03   &  0.5392 &  64 & 18.69 & 0.9 &  61.22 &  61.22 & 1 $\times$ 1 & 100.0   \\
MS0839+29   &  0.1928 &  39 & 18.70 & 0.9 & 176.35 & 167.47 & 3 $\times$ 1&  68.3  \\
MS1006+12   &  0.2605 &  29 & 18.14 & 0.9 &  59.13 &  57.51 & 1 $\times$ 1&  81.0  \\
MS1008-12   &  0.3062 &  61 & 17.89 & 1.0 &63.01 &  45.29 & 1 $\times$ 1&  73.1  \\
MS1224+20   &  0.3255 &  29 & 18.05 & 0.8 &  57.07 & 71.72  & 1 $\times$ 1 & 93.5  \\
MS1231+15   &  0.2350 &  67 & 18.66 & 0.9 & 181.35 & 181.35 & 1 $\times$ 3 & 100.0   \\
MS1358+62   &  0.3290 & 136 & 17.93 & 1.2 & 193.71 & 185.14 & 1 $\times$ 3 & 100.0   \\
MS1455+22   &  0.2570 &  57 & 18.22 & 0.9 &  58.69 &  56.39 & 1 $\times$ 1 & 77.3  \\
MS1512+36   &  0.3726 &  23 & 18.75 & 1.0 & 207.61 & 181.45 & 3 $\times$ 1 & 100.0   \\
MS1621+26   &  0.4274 &  59 & 18.85 & 1.1 &  72.44 &  67.72 & 1 $\times$ 1&  100.0   \\ 

\enddata

\tablecomments{(3) Number of spectroscopic clusters members with R $\leq$
R$_{200}$, (4) 5$\sigma$ limiting magnitude of observations, (6) Total
area with $g$ and $r$ band data, (7) Total area with both K-band and
$g$/$r$ data (9)
Percentage of a circle with radius R$_{200}$ with K-band imaging}

\end{deluxetable}
\begin{deluxetable}{lcc}
\tablecolumns{3}
\tablecaption{Comparison of 2MASS and CNOC1 Photometric Zeropoints}
\tablewidth{3.0in}
\tablehead{\colhead{Cluster} & \colhead{N$_{stars}$} &
\colhead{K$_{CNOC}$-K$_{2MASS}$} \\ 
\colhead{(1)} & \colhead{(2)}& \colhead{(3)}
}
\startdata
A2390       &  70 & 0.042 $\pm$ 0.045  \\  
MS0016+16   &  7 &  0.180 $\pm$ 0.038  \\
MS0302+16   &  9 &  0.003 $\pm$ 0.052  \\
MS0440+02   &  62 & 0.064 $\pm$ 0.015 \\
MS0451+02   &  43 & 0.005 $\pm$ 0.030  \\
MS0451-03   &  15 & 0.058 $\pm$ 0.038  \\
MS0839+29   &  27 & 0.025 $\pm$ 0.028  \\
MS1006+12   &  7 &  0.048 $\pm$ 0.058  \\
MS1008-12   &  6 &  -0.003 $\pm$ 0.080  \\
MS1224+20   &  7 &  0.009 $\pm$ 0.043  \\
MS1231+15   &  9 &  -0.068 $\pm$ 0.079   \\
MS1358+62   &  23 & -0.043 $\pm$ 0.043   \\
MS1455+22   &  6 &  0.117 $\pm$ 0.079  \\
MS1512+36   &  18 &  -0.097 $\pm$ 0.051   \\
MS1621+26   &  20 &  0.114 $\pm$ 0.084   \\ 

\enddata
\end{deluxetable}

\begin{deluxetable}{lccccc}
\tablecolumns{6}
\tablecaption{Summary of LF Parameters}
\tablewidth{5.3in}
\tablehead{\colhead{Redshift} & \colhead{Type} &
\colhead{Environment} & \colhead{K$^{*}$} &  \colhead{$\alpha$}
  \\ 
\colhead{(1)}& \colhead{(2)}& \colhead{(3)}& \colhead{(4)}&
\colhead{(5)}}
\startdata
0.296 & all & - & -24.53$\pm$0.15 & -0.84 $\pm$0.08 \\
$''$ & all & - & -24.93$\pm$0.04 & -1.1 \\
$''$ & ELL & - & -24.09$\pm$0.14 & 0.17 $\pm$ 0.18 \\
$''$ & EM+BAL & - & -24.27$\pm$0.27 & -0.95 $\pm$ 0.27 \\
\hline
0.210  &  all  &  - & 15.52$\pm$0.06 & -0.9  \\
$''$ &  ELL  &  - & 16.17$\pm$0.06 & 0.2  \\
$''$   &  EM+BAL  &  - & 16.06$\pm$0.12 & -0.9  \\
0.290  &  all  &  - & 15.93$\pm$0.06 & -0.9  \\
$''$ &  ELL  &  - & 16.72$\pm$0.07 & 0.2  \\
$''$  &  EM+BAL  &  - & 16.12$\pm$0.18 & -0.9  \\
0.462  &  all  &  - & 17.14$\pm$0.06 & -0.9  \\
$''$  &  ELL  &  - & 17.44$\pm$0.06 & 0.2  \\
$''$  &  EM+BAL  &  - & 17.20$\pm$0.14 & -0.9  \\
\hline
0.296  & all    &  Cluster-Low Mass &-24.51$\pm$0.08 & -0.9  \\  
$''$  & ELL    &  $''$ & -23.94$\pm$0.08 & 0.2  \\
$''$  & EM+BAL    &  $''$ & -23.85$\pm$0.17 & -0.9  \\
$''$  & all    &  Cluster-Mid Mass & -24.59$\pm$0.06 & -0.9 \\  
$''$  &  ELL   & $''$   & -23.99$\pm$0.05 & 0.2 \\
$''$  &  EM+BAL    &  $''$ & -24.08$\pm$0.14 & -0.9  \\
$''$  & all    &  Cluster-High Mass & -24.52$\pm$0.06 & -0.9  \\  
$''$  &  ELL    &  $''$  & -24.10$\pm$0.05 & 0.2  \\
$''$  &  EM+BAL    &  $''$  & -23.72$\pm$0.09 & -0.9  \\
\enddata
\end{deluxetable}

\end{document}